\pdfoutput=1
\documentclass[12pt,a4paper]{article}

\usepackage{ifthen} 
\newboolean{pdflatex}
\setboolean{pdflatex}{true} 

\newboolean{articletitles}
\setboolean{articletitles}{true} 

\newboolean{uprightparticles}
\setboolean{uprightparticles}{false} 

\def\paperauthors{LHCb collaboration} 
\def\paperasciititle{A model-independent measurement of the CKM angle gamma in the decays B2DhD2KKpipi and B2DhD2pipipipi} 
\def\papertitle{A model-independent measurement of the CKM angle $\gamma$ in the decays $\Bpm\to[\Kp\Km\pip\pim]_\D h^\pm$ and $\Bpm\to[\pip\pim\pip\pim]_\D h^\pm$ ($h = \kaon, \pion$)} 
\def\paperkeywords{{High Energy Physics}, {LHCb}} 
\def\papercopyright{\the\year\ CERN for the benefit of the LHCb collaboration} 
\def\paperlicence{CC BY 4.0 licence}
\def\paperlicenceurl{https://creativecommons.org/licenses/by/4.0/}

\usepackage{booktabs}

\usepackage{subcaption}

\newif\ifEnableSectionTOCLinks
\EnableSectionTOCLinksfalse 

\usepackage[top=1in, bottom=1.25in, left=1in, right=1in]{geometry}

\usepackage{microtype}
\usepackage{lineno}  
\usepackage{xspace} 
\usepackage{caption} 

\usepackage{graphicx}  
\usepackage{color}
\usepackage{colortbl}
\graphicspath{{./figs/}} 

\usepackage{amsmath} 
\usepackage{amssymb}
\usepackage{amsfonts}
\usepackage{upgreek} 

\newcommand*\patchAmsMathEnvironmentForLineno[1]{%
\expandafter\let\csname old#1\expandafter\endcsname\csname #1\endcsname
\expandafter\let\csname oldend#1\expandafter\endcsname\csname
end#1\endcsname
 \renewenvironment{#1}%
   {\linenomath\csname old#1\endcsname}%
   {\csname oldend#1\endcsname\endlinenomath}%
}
\newcommand*\patchBothAmsMathEnvironmentsForLineno[1]{%
  \patchAmsMathEnvironmentForLineno{#1}%
  \patchAmsMathEnvironmentForLineno{#1*}%
}
\AtBeginDocument{%
\patchBothAmsMathEnvironmentsForLineno{equation}%
\patchBothAmsMathEnvironmentsForLineno{align}%
\patchBothAmsMathEnvironmentsForLineno{flalign}%
\patchBothAmsMathEnvironmentsForLineno{alignat}%
\patchBothAmsMathEnvironmentsForLineno{gather}%
\patchBothAmsMathEnvironmentsForLineno{multline}%
\patchBothAmsMathEnvironmentsForLineno{eqnarray}%
}

\usepackage{hyperxmp}

\usepackage[pdftex,
            pdfauthor={\paperauthors},
            pdftitle={\paperasciititle},
            pdfkeywords={\paperkeywords},
            pdfcopyright={Copyright (C) \papercopyright},
            pdflicenseurl={\paperlicenceurl}]{hyperref}

\usepackage[bottom,flushmargin,hang,multiple]{footmisc}

\usepackage[all]{hypcap} 

\usepackage{xspace} 
\usepackage{upgreek}

\def\lhcb   {\mbox{LHCb}\xspace}

\def\besiii {\mbox{BESIII}\xspace}
\def\cleo   {\mbox{CLEO}\xspace}

\def\MagUp {\mbox{\em Mag\kern -0.05em Up}\xspace}

\ifthenelse{\boolean{uprightparticles}}%
{

 \def\Ppi         {\ensuremath{\uppi}\xspace}

 \def\Ppsi        {\ensuremath{\uppsi}\xspace}

 \def\PDelta      {\ensuremath{\Delta}\xspace}                 
 \def\PXi         {\ensuremath{\Xi}\xspace}                 
 \def\PLambda     {\ensuremath{\Lambda}\xspace}                 
 \def\PSigma      {\ensuremath{\Sigma}\xspace}                 
 \def\POmega      {\ensuremath{\Omega}\xspace}                 
 \def\PUpsilon    {\ensuremath{\Upsilon}\xspace}
 \let\oldPi\Pi
 \def\PPi         {\ensuremath{\oldPi}\xspace}

 \def\PB      {\ensuremath{\mathrm{B}}\xspace}                 
 \def\PD      {\ensuremath{\mathrm{D}}\xspace}                 

 \def\PK      {\ensuremath{\mathrm{K}}\xspace}                 
 \def\Pb      {\ensuremath{\mathrm{b}}\xspace}                 
 \def\Pc      {\ensuremath{\mathrm{c}}\xspace}                 
 \def\Pd      {\ensuremath{\mathrm{d}}\xspace}                 
 \def\Pe      {\ensuremath{\mathrm{e}}\xspace}                 
 \def\Ps      {\ensuremath{\mathrm{s}}\xspace}                 
                  
 \def\Pu      {\ensuremath{\mathrm{u}}\xspace}                 
 \def\thebaroffset{0.0em}
}
{

 \def\Ppi         {\ensuremath{\pi}\xspace}

 \def\Ppsi        {\ensuremath{\psi}\xspace}                 
                  
 \mathchardef\PDelta="7101
 \mathchardef\PXi="7104
 \mathchardef\PLambda="7103
 \mathchardef\PSigma="7106
 \mathchardef\POmega="710A
 \mathchardef\PUpsilon="7107
 \mathchardef\PPi="7105
 \def\PB      {\ensuremath{B}\xspace}                 
 \def\PD      {\ensuremath{D}\xspace}                 

 \def\PK      {\ensuremath{K}\xspace}                 
 \def\Pb      {\ensuremath{b}\xspace}                 
 \def\Pc      {\ensuremath{c}\xspace}                 
 \def\Pd      {\ensuremath{d}\xspace}                 
 \def\Pe      {\ensuremath{e}\xspace}                 
 \def\Ps      {\ensuremath{s}\xspace}                 
                  
 \def\Pu      {\ensuremath{u}\xspace}                 
 \def\thebaroffset{0.18em}
}
\newcommand{\offsetoverline}[2][\thebaroffset]{\kern #1\overline{\kern -#1 #2}}%

\makeatletter
\ifcase \@ptsize \relax
  \newcommand{\miniscule}{\@setfontsize\miniscule{4}{5}}
\or
  \newcommand{\miniscule}{\@setfontsize\miniscule{5}{6}}
\or
  \newcommand{\miniscule}{\@setfontsize\miniscule{5}{6}}
\fi
\makeatother

\DeclareRobustCommand{\optbar}[1]{\shortstack{{\miniscule (\rule[.5ex]{1.25em}{.18mm})}
  \\ [-.7ex] $#1$}}

\def\en         {{\ensuremath{\Pe^-}}\xspace}   
\def\ep         {{\ensuremath{\Pe^+}}\xspace}

\def\uquark    {{\ensuremath{\Pu}}\xspace}
\def\uquarkbar {{\ensuremath{\overline \uquark}}\xspace}

\def\dquark    {{\ensuremath{\Pd}}\xspace}

\def\squark    {{\ensuremath{\Ps}}\xspace}

\def\cquark    {{\ensuremath{\Pc}}\xspace}
\def\cquarkbar {{\ensuremath{\overline \cquark}}\xspace}

\def\bquark    {{\ensuremath{\Pb}}\xspace}

\def\pion   {{\ensuremath{\Ppi}}\xspace}
\def\piz    {{\ensuremath{\pion^0}}\xspace}
\def\pip    {{\ensuremath{\pion^+}}\xspace}
\def\pim    {{\ensuremath{\pion^-}}\xspace}
\def\pipm   {{\ensuremath{\pion^\pm}}\xspace}
\def\pimp   {{\ensuremath{\pion^\mp}}\xspace}

\def\kaon    {{\ensuremath{\PK}}\xspace}

\def\KorKbar {\kern \thebaroffset\optbar{\kern -\thebaroffset \PK}{}\xspace}

\def\Kp      {{\ensuremath{\kaon^+}}\xspace}
\def\Km      {{\ensuremath{\kaon^-}}\xspace}
\def\Kpm     {{\ensuremath{\kaon^\pm}}\xspace}
\def\Kmp     {{\ensuremath{\kaon^\mp}}\xspace}
\def\KS      {{\ensuremath{\kaon^0_{\mathrm{S}}}}\xspace}

\def\Dbar    {{\ensuremath{\offsetoverline{\PD}}}\xspace}
\def\D       {{\ensuremath{\PD}}\xspace}
\def\Db      {{\ensuremath{\Dbar}}\xspace}
\def\DorDbar {\kern \thebaroffset\optbar{\kern -\thebaroffset \PD}\xspace}
\def\Dz      {{\ensuremath{\D^0}}\xspace}
\def\Dzb     {{\ensuremath{\Dbar{}^0}}\xspace}
\def\Dp      {{\ensuremath{\D^+}}\xspace}
\def\Dm      {{\ensuremath{\D^-}}\xspace}

\def\DpDm    {\ensuremath{\Dp {\kern -0.16em \Dm}}\xspace}
\def\Dstar   {{\ensuremath{\D^*}}\xspace}

\def\B       {{\ensuremath{\PB}}\xspace}
\def\Bbar    {{\ensuremath{\offsetoverline{\PB}}}\xspace}

\def\BorBbar {\kern \thebaroffset\optbar{\kern -\thebaroffset \PB}\xspace}
\def\Bz      {{\ensuremath{\B^0}}\xspace}

\def\Bd      {{\ensuremath{\B^0}}\xspace}

\def\BdorBdbar {\kern \thebaroffset\optbar{\kern -\thebaroffset \Bd}\xspace}
\def\Bu      {{\ensuremath{\B^+}}\xspace}
\def\Bub     {{\ensuremath{\B^-}}\xspace}
\def\Bp      {{\ensuremath{\Bu}}\xspace}
\def\Bm      {{\ensuremath{\Bub}}\xspace}
\def\Bpm     {{\ensuremath{\B^\pm}}\xspace}

\def\Bs      {{\ensuremath{\B^0_\squark}}\xspace}
\def\Bsb     {{\ensuremath{\Bbar{}^0_\squark}}\xspace}
\def\BsorBsbar {\kern \thebaroffset\optbar{\kern -\thebaroffset \Bs}\xspace}

\def\psiprpr  {{\ensuremath{\Ppsi(3770)}}\xspace}

\def\Y#1S{\ensuremath{\PUpsilon{(#1S)}}\xspace}

\def\Lz          {{\ensuremath{\PLambda}}\xspace}

\def\LorLbar     {\kern \thebaroffset\optbar{\kern -\thebaroffset \PLambda}\xspace}

\def\Lb           {{\ensuremath{\Lz^0_\bquark}}\xspace}

\def\to                 {\ensuremath{\rightarrow}\xspace}

\def\CP                {{\ensuremath{C\!P}}\xspace}

\def\Vud  {{\ensuremath{V_{\uquark\dquark}^{\phantom{\ast}}}}\xspace}
\def\Vcd  {{\ensuremath{V_{\cquark\dquark}^{\phantom{\ast}}}}\xspace}

\def\Vubs  {{\ensuremath{V_{\uquark\bquark}^\ast}}\xspace}
\def\Vcbs  {{\ensuremath{V_{\cquark\bquark}^\ast}}\xspace}

\def\AT#1     {\ensuremath{A_{\mathrm{T}}^{#1}}\xspace}           

\def\C#1      {\ensuremath{\mathcal{C}_{#1}}\xspace}                       
\def\Cp#1     {\ensuremath{\mathcal{C}_{#1}^{'}}\xspace}                    
\def\Ceff#1   {\ensuremath{\mathcal{C}_{#1}^{\mathrm{(eff)}}}\xspace}        
\def\Cpeff#1  {\ensuremath{\mathcal{C}_{#1}^{'\mathrm{(eff)}}}\xspace}       
\def\Ope#1    {\ensuremath{\mathcal{O}_{#1}}\xspace}                       
\def\Opep#1   {\ensuremath{\mathcal{O}_{#1}^{'}}\xspace}                    

\newcommand{\nospaceunit}[1]{\ensuremath{\text{#1}}}       
\newcommand{\aunit}[1]{\ensuremath{\text{\,#1}}}       

\newcommand{\tev}{\aunit{Te\kern -0.1em V}\xspace}
\newcommand{\gev}{\aunit{Ge\kern -0.1em V}\xspace}
\newcommand{\mev}{\aunit{Me\kern -0.1em V}\xspace}
\newcommand{\kev}{\aunit{ke\kern -0.1em V}\xspace}
\newcommand{\ev}{\aunit{e\kern -0.1em V}\xspace}
 
\newcommand{\mevc}{\ensuremath{\aunit{Me\kern -0.1em V\!/}c}\xspace}
\newcommand{\gevc}{\ensuremath{\aunit{Ge\kern -0.1em V\!/}c}\xspace}
\newcommand{\mevcc}{\ensuremath{\aunit{Me\kern -0.1em V\!/}c^2}\xspace}
\newcommand{\gevcc}{\ensuremath{\aunit{Ge\kern -0.1em V\!/}c^2}\xspace}

\def\mum  {\ensuremath{\,\upmu\nospaceunit{m}}\xspace}

\def\fb   {\ensuremath{\aunit{fb}}\xspace}
\def\invfb   {\ensuremath{\fb^{-1}}\xspace}

\def\gsim{{~\raise.15em\hbox{$>$}\kern-.85em
          \lower.35em\hbox{$\sim$}~}\xspace}
\def\lsim{{~\raise.15em\hbox{$<$}\kern-.85em
          \lower.35em\hbox{$\sim$}~}\xspace}

\newcommand{\Real}{\ensuremath{\mathcal{R}e}\xspace}
\newcommand{\Imag}{\ensuremath{\mathcal{I}m}\xspace}

\def\pt         {\ensuremath{p_{\mathrm{T}}}\xspace}

\def\ptot       {\ensuremath{p}\xspace}

\def\evtgen     {\mbox{\textsc{EvtGen}}\xspace}

\def\geant      {\mbox{\textsc{Geant4}}\xspace}

\def\photos     {\mbox{\textsc{Photos}}\xspace}

\def\pythia     {\mbox{\textsc{Pythia}}\xspace}

\def\tell1  {TELL1\xspace}
\def\ukl1   {UKL1\xspace}

\newcommand{\lhcborcid}[1]{\href{https://orcid.org/#1}{\hspace*{0.1em}\raisebox{-0.45ex}{\includegraphics[width=1em]{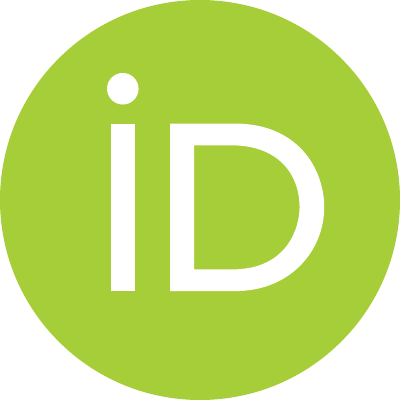}}}}

\hypersetup{
  colorlinks   = true, 
  urlcolor     = blue, 
  linkcolor    = blue, 
  citecolor    = red   
}

\ifEnableSectionTOCLinks
    \usepackage[explicit]{titlesec} 
    
    \let\oldcontentsline\contentsline
    \renewcommand

    \titleformat{\section}{\normalfont\Large\bf}{\hyperlink{tocsection.\thesection}{{\thesection} \parbox[t]{\dimexpr\textwidth-1pc}{#1}}}{1pc}{}

    \titleformat{\subsection}{\normalfont\bf}{\hyperlink{tocsubsection.\thesubsection}{{\thesubsection} \parbox[t]{\dimexpr\textwidth-1pc}{#1}}}{1pc}{}

    \titleformat{name=\section,numberless}[display]{}{}{0pt}{\normalfont\Huge\bfseries #1}
\fi

\usepackage{cite} 
\usepackage{mciteplus}

\begin{document}

\renewcommand{\thefootnote}{\fnsymbol{footnote}}
\setcounter{footnote}{1}


\begin{titlepage}
\pagenumbering{roman}

\vspace*{-1.5cm}
\centerline{\large EUROPEAN ORGANIZATION FOR NUCLEAR RESEARCH (CERN)}
\vspace*{1.5cm}
\noindent
\begin{tabular*}{\linewidth}{lc@{\extracolsep{\fill}}r@{\extracolsep{0pt}}}
\ifthenelse{\boolean{pdflatex}}
{\vspace*{-1.5cm}\mbox{\!\!\!\includegraphics[width=.14\textwidth]{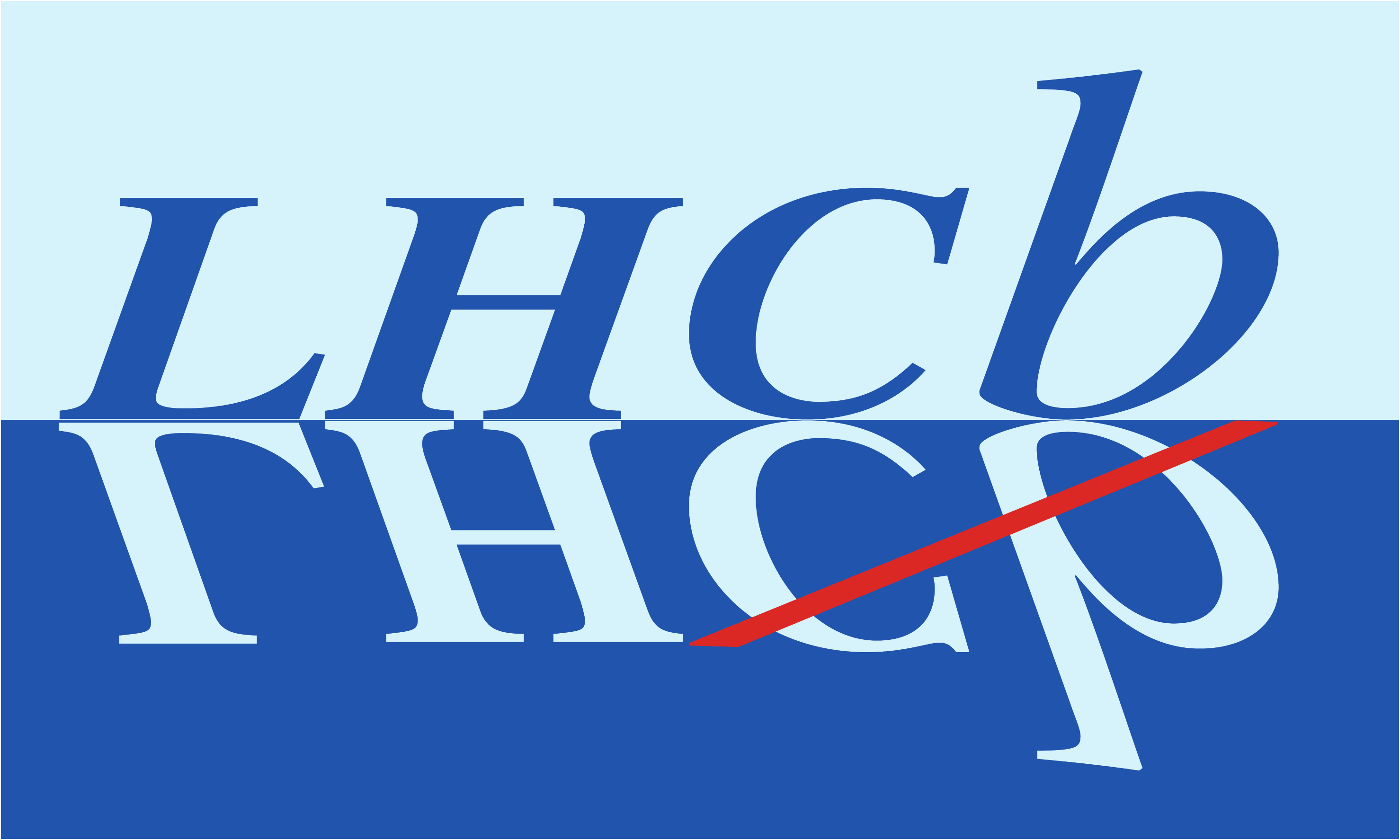}} & &}%
{\vspace*{-1.2cm}\mbox{\!\!\!\includegraphics[width=.12\textwidth]{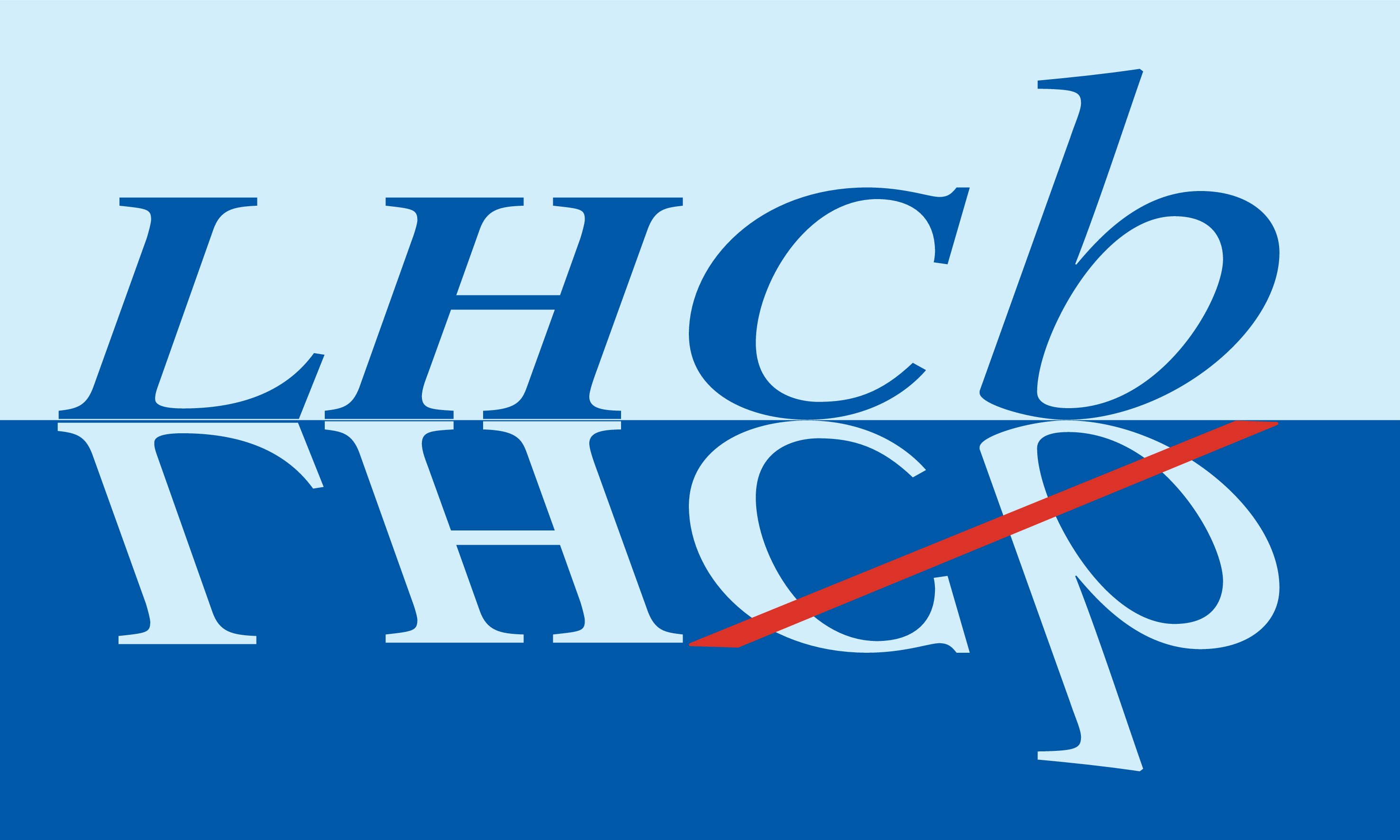}} & &}%
\\
 & & CERN-EP-2025-199 \\  
 & & LHCb-PAPER-2025-019 \\  
 & & January 9, 2026 \\ 
 & & \\
\end{tabular*}

\vspace*{4.0cm}

{\normalfont\bfseries\boldmath\huge
\begin{center}
  \papertitle 
\end{center}
}

\vspace*{2.0cm}

\begin{center}
\paperauthors\footnote{Authors are listed at the end of this paper.}
\end{center}

\vspace{\fill}

\begin{abstract}
  \noindent
  A model-independent determination of the CKM angle $\gamma$ is presented, using the $\Bpm\to[\Kp\Km\pip\pim]_\D h^\pm$ and $\Bpm\to[\pip\pim\pip\pim]_\D h^\pm$ decays, with $h=K,\pi$. This measurement is the first phase-space binned study of these decay modes, and uses a sample of proton-proton collision data collected by the \lhcb experiment, corresponding to an integrated luminosity of $9\invfb$. The phase-space bins are optimised for sensitivity to $\gamma$, and in each bin external inputs from the \besiii experiment are used to constrain the charm strong-phase parameters. The result of this binned analysis is $\gamma = (53.9_{-8.9}^{+9.5})^\circ$, where the uncertainty includes both statistical and systematic contributions. Furthermore, when combining with existing phase-space integrated measurements of the same decay modes, a value of  $\gamma = (52.6_{-6.4}^{+8.5})^\circ$ is obtained, which is one of the most precise determinations of $\gamma$ to date.
\end{abstract}

\vspace*{1.49cm}

\begin{center}
  Published in JHEP 01 (2026) 062
\end{center}

\vspace{\fill}

{\footnotesize 
\centerline{\copyright~\papercopyright. \href{\paperlicenceurl}{\paperlicence}.}}
\vspace*{2mm}

\end{titlepage}


\newpage
\setcounter{page}{2}
\mbox{~}
%
%
%
%


\renewcommand{\thefootnote}{\arabic{footnote}}
\setcounter{footnote}{0}


\cleardoublepage


\pagestyle{plain} 
\setcounter{page}{1}
\pagenumbering{arabic}


\section{Introduction}
\label{section:Introduction}
A long-standing open question in physics concerns the nature and origin of the \CP violation necessary to explain the matter-antimatter asymmetry present in the Universe~\cite{cite:Sakharov}. In the Standard Model (SM) of particle physics, \CP violation in the quark sector is described by the Cabibbo--Kobayashi--Maskawa (CKM) matrix~\cite{Cabibbo:1963yz,Kobayashi:1973fv}. Specifically, this matrix contains a complex irreducible phase that is responsible for all \CP violation observed to date in particle interactions. However, the size of \CP-violating effects seen in the quark sector is not sufficient to explain the matter-antimatter asymmetry of the Universe as a whole. Therefore, there could be other potential sources of \CP violation beyond the SM.

The unitary nature of the CKM matrix leads to a set of constraints between its elements, which can be visualised as closed triangles in the Argand plane. One such triangle, known as the Unitarity Triangle, is of particular interest. The lengths and angles of this triangle can be experimentally determined from properties of $\B$ decays, and from these measurements it is possible to probe for discrepancies that could give hints of \CP violation and other effects arising from New Physics (NP).

The CKM angle $\gamma$, defined as $\gamma\equiv\arg(-\Vud\Vubs/\Vcd\Vcbs)$, is the only CKM angle that can be measured directly in tree-level decays, with negligible theoretical uncertainties~\cite{cite:Brod_Zupan}. Such measurements are therefore valuable as a standard candle of the SM, as they may be compared with other indirect loop-level determinations of $\gamma$ that can be susceptible to NP effects beyond the SM, assuming there are no NP effects in tree-level decays~\cite{Brod:2014bfa,Lenz:2019lvd}. Two such indirect determinations have been performed in Refs.~\cite{CKMfitter2015} and \cite{cite:UTfit}, with the results $\gamma = (66.3_{-1.9}^{+0.7})^\circ$ and $\gamma = (64.9\pm1.4)^\circ$, respectively. These loop-level determinations are mostly driven by measurements of the CKM angle $\beta$, the mixing frequencies $\Delta m_d$ and $\Delta m_s$ of $\Bz$ and $\Bs$ mesons, respectively, and inputs from lattice QCD.

A direct measurement of $\gamma$, which requires interference between $\bquark\to \cquark\uquarkbar\squark$ and $\bquark\to\uquark\cquarkbar\squark$ quark transitions, can be performed using the decay channel $\Bpm\to\D\Kpm$. In this expression, $\D$ represents a superposition of the neutral charm mesons $\Dz$ and $\Dzb$, which arises since the $\Bm\to\D\Km$ decay process can proceed through both the favoured $\Bm\to\Dz\Km$ and the suppressed $\Bm\to\Dzb\Km$ decay. In the charge-conjugate decay, $\Bp\to\D\Kp$, an analogous superposition of $\Dz$ and $\Dzb$ mesons occurs. The $\Dz$ and $\Dzb$ must subsequently decay into a common final state in order for interference between the corresponding amplitudes to occur.

In addition, similar $\bquark$-hadron decay modes, such as $\Bpm\to\Dstar h^\pm$, where $h = \pi$ or $K$, may also be exploited to obtain the best possible statistical precision. A combination of all such measurements by the \lhcb collaboration to date results in a value of $\gamma = (64.4 \pm 2.8)^\circ$~\cite{LHCb-CONF-2024-004}, which, though consistent with the loop-level determination, is significantly less precise. Hence, it is important to exploit new decay modes and employ new analysis methods to obtain improved knowledge of $\gamma$ through tree-level processes.

A powerful class of neutral charm decays that can be used in the determination of $\gamma$ is that of multibody final states of mixed \CP content. In these transitions, the amplitude of the $\Dz$ and $\Dzb$ decays have strong-phase differences that vary across the phase space. Although a phase-space integrated measurement of $\gamma$ would have limited sensitivity due to the dilution of interference effects, a binned measurement, where phase-space regions of similar strong-phase difference are grouped together, can achieve a unique solution for $\gamma$ with high precision. To date, the most precise binned measurement is $\gamma = (68.7_{-5.1}^{+5.2})^\circ$ using $\D\to\KS h^+h^-$ decays, which is reported in Ref.~\cite{LHCb-PAPER-2020-019}. The analysis takes as input external measurements of the amplitude-averaged charm strong-phase differences in each phase-space bin. These are obtained from quantum-correlated $\D\Db$ pairs produced at the $\psiprpr$ threshold~\cite{cite:KShh_cisi_CLEO,cite:KSpipi_cisi_PRL,cite:KSpipi_cisi_PRD,cite:KSKK_cisi}.

An analogous extension of this approach can be applied to four-body charm decays~\cite{Rademacker:2006zx}. In Ref.~\cite{LHCb-PAPER-2022-037}, it has been demonstrated that an optimal binning scheme for the decay $\D\to\Kp\Km\pip\pim$ can achieve competitive statistical precision with other decay modes that contribute to the combination of $\gamma$ measurements by the \lhcb collaboration~\cite{LHCb-CONF-2024-004}. The binning scheme was constructed using knowledge from an amplitude model~\cite{LHCb-PAPER-2018-041}. At the time of this previous $\gamma$ measurement, no strong-phase information was available. Consequently, the same amplitude model was used to calculate the strong-phase inputs. This model dependence introduced systematic uncertainties that are challenging to evaluate.

In this paper, a model-independent determination of $\gamma$ using $\Bpm\to\D\Kpm$, with the subsequent $\D\to\Kp\Km\pip\pim$ decay, is presented. The analysis is performed in bins of phase space and takes as input direct measurements of the strong-phase parameters recently reported by the \besiii collaboration~\cite{cite:KKpipi_cisi_BESIII}. This is the first binned model-independent measurement of $\gamma$ exploiting this four-body $\D$-decay mode, and supersedes the model-dependent measurement in Ref.~\cite{LHCb-PAPER-2022-037}. 

Analogously, the decay $\D\to\pip\pim\pip\pim$ may also be used in a measurement of $\gamma$. This decay mode has a branching fraction approximately three times that of $\D\to\Kp\Km\pip\pim$, with similar kinematics, and it is therefore expected to also provide valuable sensitivity to $\gamma$. Data collected by the \cleo-c experiment has been used to construct an amplitude model, devise a binning scheme for the measurement of $\gamma$, and determine the strong-phase parameters within these bins~\cite{cite:4pi_cisi_CLEOc}. More recently, an amplitude model, binning scheme and strong-phase measurements for this same decay mode have been reported by the \besiii collaboration~\cite{cite:4pi_model_BESIII,cite:4pi_cisi_BESIII}. As the \besiii data sample is significantly larger than that accumulated by \cleo-c, it is expected that the \besiii choice of binning allows for a better sensitivity to $\gamma$, as does the higher precision of the strong-phase measurements within these bins. This paper presents a first determination of $\gamma$ that exploits these \besiii measurements of $\D\to\pip\pim\pip\pim$ strong-phase differences, and also a cross-check using the earlier \cleo-c inputs. Finally, a measurement of $\gamma$ is reported that combines the results from both the $\D\to\Kp\Km\pip\pim$ and $\D\to\pip\pim\pip\pim$ decay modes.

\section{Analysis strategy}
\label{section:Analysis_strategy}
The formalism for measuring the CKM angle $\gamma$ in this analysis is presented in Ref.~\cite{LHCb-PAPER-2022-037}. In summary, the analysis of multibody decays with mixed \CP-content can be performed with a binned phase-space approach, which is the main focus of this paper. This requires a determination of the yield $N_i^\pm$ of $\Bpm$ decays in the $2\times\mathcal{N}$ bins of the $\D$-meson phase space, labelled $i = -\mathcal{N}, ..., -1, +1, ..., \mathcal{N}$. The bins where $i < 0$ are found by \CP-conjugating the corresponding phase-space regions labelled by a positive bin number. The key equations relating the yields $N_i^\pm$ of $\Bpm$ decays in bin $i$ are given by
\begin{align}
  N_{+i}^+ =& h_\Bp^{\D\kaon}\Big(F_{-i} + \big((x_+^{\D\kaon})^2 + (y_+^{\D\kaon})^2\big)F_i + 2\sqrt{F_{-i}F_i}(x_+^{\D\kaon}c_i - y_+^{\D\kaon}s_i)\Big), \label{equation:Bp_yield}  \\
  N_{-i}^- =& h_\Bm^{\D\kaon}\Big(F_{-i} + \big((x_-^{\D\kaon})^2 + (y_-^{\D\kaon})^2\big)F_i + 2\sqrt{F_{-i}F_i}(x_-^{\D\kaon}c_i - y_-^{\D\kaon}s_i)\Big), \label{equation:Bm_yield}
\end{align}
where the \CP-violating observables $x_{\pm}^{\D\kaon}$ and $y_{\pm}^{\D\kaon}$ are related to $\gamma$ by
\begin{equation}
  x_\pm^{\D\kaon} \equiv r_\B^{\D\kaon}\cos(\delta_\B^{\D\kaon} \pm \gamma), \quad y_\pm^{\D\kaon} \equiv r_\B^{\D\kaon}\sin(\delta_\B^{\D\kaon} \pm \gamma).
  \label{equation:Cartesian_parametersiation}
\end{equation}
 In Eq.~\eqref{equation:Cartesian_parametersiation}, the parameters $r_\B^{\D\kaon}$ and $\delta_\B^{\D\kaon}$ are, respectively, the ratio of the magnitude and the strong-phase difference between the suppressed $\Bm\to\Dzb\Km$ and favoured $\Bm\to\Dz\Km$ decay amplitudes.

The amplitude-averaged cosine and sine of the strong phase of the charm decay within each bin, $c_i$ and $s_i$, are provided as external inputs for $i > 0$. For the \CP-conjugated bins, it can be shown that these parameters satisfy $c_{-i} = c_i$ and $s_{-i} = -s_i$ if \CP-violating effects are neglected for the charm-meson decay~\cite{cite:CharmCPVgamma}.

The parameters $F_i$ describe the fraction of $\Dz$ decays in bin $i$, in the absence of interference effects. Moreover, the $F_i$ parameters are treated as free parameters, which allow them to absorb effects such as nonuniform phase-space acceptance, charm mixing and bin migration, making the analysis insensitive to such effects. The normalisation parameters $h_\Bp^{\D\kaon}$ and $h_\Bm^{\D\kaon}$ are kept independent between the $\Bp$ and $\Bm$ decays to ensure that the measurement is insensitive to any global production or detection asymmetries, as well as global \CP-violation effects.

The mode $\Bpm\to\D\pipm$ is kinematically similar to the $\Bpm\to\D\Kpm$ decay, but with a branching fraction that is an order of magnitude larger~\cite{PDG2024}. Furthermore, while \CP-violating effects in the latter mode are around $10\%$, as determined by the magnitude of the interference parameter $r_\B^{\D\kaon} \approx 0.10$, the $\Bpm\to\D\pipm$ decay has very minor interference effects, with $r_\B^{\D\pion} \approx 0.005$~\cite{LHCb-CONF-2024-004}. This makes this decay mode an ideal normalisation channel, and the parameters $F_i$, $c_i$ and $s_i$ are shared between the two $\Bpm$-decay modes. However, the small \CP-violating effects in the $\Bpm\to\D\pipm$ decay, which depend on $\gamma$, could potentially bias its determination if such effects are not properly described by the fit model. Therefore, the $\Bpm\to\D\Kpm$ and $\Bpm\to\D\pipm$ channels are fitted simultaneously, and \CP-violating observables are extracted for both. The equations for $\Bpm\to\D\pipm$ are analogous to that of Eqs.~\eqref{equation:Bp_yield} and \eqref{equation:Bm_yield}, but with the $\D\kaon$ superscript replaced with $\D\pion$. The advantage of this setup is that the $\Bpm\to\D\pipm$ channel mostly constrains the $F_i$ parameters, while most of the sensitivity to $\gamma$ is provided by the $\Bpm\to\D\Kpm$ mode.

In their current form, Eqs.~\eqref{equation:Bp_yield} and \eqref{equation:Bm_yield} contain several redundancies. First, when including the $\Bpm\to\D\pipm$ mode, there are eight \CP-violating observables in total, but they correspond to only five parameters $\gamma$, $\delta_\B^{\D\kaon}$, $r_\B^{\D\kaon}$, $\delta_\B^{\D\pion}$ and $r_\B^{\D\pion}$, which are referred to as physics parameters. A more convenient parameterisation for the $\Bpm\to\D\pipm$ mode is~\cite{cite:Simultaneous_B_gamma_fit}
\begin{equation}
  x_\xi^{\D\pion} \equiv \Real(\xi^{\D\pion}), \quad y_\xi^{\D\pion} \equiv \Imag(\xi^{\D\pion}),\quad \xi^{\D\pion} \equiv \frac{r_\B^{\D\pion}}{r_\B^{\D\kaon}}\exp\Big(i(\delta_\B^{\D\pion} - \delta_\B^{\D\kaon})\Big),
  \label{equation:xi}
\end{equation}
such that the new set of six \CP-violating observables consists of $x_\pm^{\D\kaon}$, $y_\pm^{\D\kaon}$, $x_\xi^{\D\pion}$ and $y_\xi^{\D\pion}$.

The second redundancy is due to the definition of $F_i$, which constrains $\sum_iF_i = 1$. An alternative parameterisation, using the recursive fractions defined by 
\begin{equation}
    F_i \equiv
    \begin{cases}
        R_i, & i = -\mathcal{N} \\
        R_i\prod_{j < i}(1 - R_j), & -\mathcal{N} < i < +\mathcal{N} \\
        \prod_{j < i}(1 - R_j), &i = +\mathcal{N},
    \end{cases}
\end{equation}
trivially fixes the last parameter to $R_{\mathcal{N}} = 1$ and avoids fit instabilities due to this redundancy.

An alternative approach to measuring $\gamma$ that is complementary to the binned method, is to exploit the information contained in the phase-space integrated yields of $\Bpm$ decays. The \CP-violating observables sensitive to $\gamma$ in the $\D\to\Kp\Km\pip\pim$ channel are the global asymmetry,
\begin{equation}
    A^{KK\pi\pi}_h \equiv \frac{\Gamma(\Bm\to\D h^-) - \Gamma(\Bp\to\D h^+)}{\Gamma(\Bm\to\D h^-) + \Gamma(\Bp\to\D h^+)},
    \label{equation:Asymmetry}
\end{equation}
and the double ratio
\begin{equation}
    R^{KK\pi\pi}_{\CP} \equiv \frac{R_{\kaon\kaon\pion\pion}}{R_{K\pion\pion\pion}}, \quad R_f \equiv \frac{\Gamma(\Bm\to[f]_\D\Km) + \Gamma(\Bp\to[f]_\D\Kp)}{\Gamma(\Bm\to[f]_\D\pim) + \Gamma(\Bp\to[f]_\D\pip)},
    \label{equation:R_CP}
\end{equation}
where in the case of $R_{K\pion\pion\pion}$ the kaon from the $D$-meson decay has the same charge as the pion or kaon from the $B$-meson decay. The value of $R_{K\pion\pion\pion}$ can be determined from the phase-space integrated yields obtained in Ref.~\cite{LHCb-PAPER-2022-017}.

The relationship between these phase-space integrated \CP-violating observables and the underlying physics parameters may be expressed as~\cite{LHCb-PAPER-2022-037},
\begin{align}
    A^{KK\pi\pi}_h &= \frac{2r_\B^{\D h}\kappa\sin(\delta_\B^{\D h})\sin(\gamma)}{1 + (r_\B^{\D h})^2 + 2r_\B^{\D h}\kappa\cos(\delta_\B^{\D h})\cos(\gamma)}, \label{equation:Asymmetry_gamma} \\[0.5em]
    R^{KK\pi\pi}_{\CP} &= 1 + (r_\B^{\D\kaon})^2 + 2r_\B^{\D\kaon}\kappa\cos(\delta_\B^{\D\kaon})\cos(\gamma), \label{equation:R_CP_gamma}
\end{align}
where $\kappa = 2F^{KK\pi\pi}_+ - 1$ is the dilution factor when integrating over the full phase space, and $F^{KK\pi\pi}_+$ is the \CP-even fraction of the decay. Analogous observables $A^{\pi\pi\pi\pi}_h$ and $R^{\pi\pi\pi\pi}_{\CP}$ exist for the $D \to \pip\pim\pip\pim$ decay.

Measurements of the asymmetries $A_\kaon$, $A_\pion$ and the double ratio $R_{\CP}$ are reported for the $\D\to\Kp\Km\pip\pim$ and $\pip\pim\pip\pim$ channels in Ref.~\cite{LHCb-PAPER-2022-037}. Their correlations with the binned \CP-violating observables are found to be negligible, since the normalisation parameters $h_\Bp^{\D\kaon}$ and $h_\Bm^{\D\kaon}$ that contain information about global asymmetries are not used in the subsequent interpretation of binned \CP-violating observables in terms of $\gamma$.

Thus, to provide the maximum sensitivity to $\gamma$ that can be achieved with these four-body decay modes, the phase-space binned \CP-violating observables measured in this analysis are combined with the phase-space integrated observables from Ref.~\cite{LHCb-PAPER-2022-037}. In the combination, Eqs.~\eqref{equation:Asymmetry_gamma} and \eqref{equation:R_CP_gamma} are corrected for small effects of charm mixing, using the formalism from Ref.~\cite{Rama:2013voa} and mixing parameters from Ref.~\cite{HFLAV21}.

\section{Binning schemes and external-charm inputs}
\label{section:Binning_schemes_and_external_charm_inputs}
The analysis strategy described in Sect.~\ref{section:Analysis_strategy} requires a binning scheme, which is a set of instructions for assigning $\D$-decay candidates a phase-space bin. In an $N$-body decay, the phase-space dimensionality is $3N - 7$ for $N\geq3$. For three-body decays, such as $\D\to\KS\pip\pim$, the binning scheme may be represented on a Dalitz plane, as demonstrated in Refs.~\cite{cite:KShh_cisi_CLEO,cite:KSpipi_cisi_PRL,cite:KSpipi_cisi_PRD}. The binning scheme for a four-body decay, where the phase space is five-dimensional, is unfortunately not easily visualised. Nevertheless, several attempts to develop such five-dimensional binning schemes have been performed in the past, such as those for the $\D\to\KS\pip\pim\piz$~\cite{cite:KSpipipi0_cisi} and $\D\to\Kpm\pimp\pipm\pimp$~\cite{cite:Kpipipi_deltaD} decays.

To create a binning scheme that has optimal sensitivity to $\gamma$, it is advantageous to use an amplitude model. This allows for a good understanding of variations in the strong-phase difference between $\Dz$ and $\Dzb$ decays throughout the phase space. In addition, an amplitude model is convenient for assessing the performance of a model-independent binned fit, relative to a model-dependent unbinned fit. In the $\D\to\Kp\Km\pip\pim$ decay, a set of binning schemes was introduced in Ref.~\cite{LHCb-PAPER-2022-037} with $2\times4$ and $2\times8$ bins. The bin number assigned to each $\D$-decay candidate is based solely on the predicted strong-phase difference and the ratio of amplitudes of the $\Dz$ and $\Dzb$ decays to this final state. The prediction is calculated using the model presented in Ref.~\cite{LHCb-PAPER-2018-041}, which takes the four-momenta of the four $\D$-decay products as input. The implementation of these binning schemes can be found in Ref.~\cite{cite:KKpipiBinningScheme}.

The strong phase parameters $c_i$ and $s_i$ have been measured in the $2\times4$ binning scheme in Ref.~\cite{cite:KKpipi_cisi_BESIII} and are presented in Table~\ref{table:cisi_KKpipi}. Their correlations are provided in Ref.~\cite{cite:KKpipi_cisi_BESIII}. The measurement was performed using data from $\ep\en$ collisions collected at the $\D\Db$ threshold by the \besiii experiment, corresponding to an integrated luminosity of $20\invfb$.

In the $\D\to\pip\pim\pip\pim$ channel, the binning scheme formalism is described in Ref.~\cite{cite:4pi_cisi_CLEOc}. Similar to the $\D\to\Kp\Km\pip\pim$ decay, the bin-number assignment of each $\D\to\pip\pim\pip\pim$ candidate is guided by an amplitude model fitted to CLEO-c data~\cite{cite:4pi_model_CLEOc}. The implementation uses a binary tree structure to form an efficient lookup table across five dimensions. In Ref.~\cite{cite:4pi_cisi_CLEOc}, binning schemes with $2\times\mathcal{N}$ bins are provided in the supplementary material, with $\mathcal{N}$ ranging from one to five.

Recently, a new amplitude model describing the $\D\to\pip\pim\pip\pim$ decay has become available from the BESIII collaboration~\cite{cite:4pi_model_BESIII}. Using the software framework from Ref.~\cite{cite:4pi_cisi_CLEOc}, an improved binning scheme with $2\times5$ bins has been constructed~\cite{cite:4pi_cisi_BESIII,cite:4piBinningScheme}. The new measurements of $c_i$ and $s_i$ from Ref.~\cite{cite:4pi_cisi_BESIII}, measured using \besiii data corresponding to an integrated luminosity of $3\invfb$, are shown in Table~\ref{table:cisi_4pi}. The correlation matrix can be found in Ref.~\cite{cite:4pi_cisi_BESIII}.

\begin{table}[!b]
  \caption{Measured values of $c_i$ and $s_i$ for the $\D\to\Kp\Km\pip\pim$ decay that are used in this analysis, taken from Ref.~\cite{cite:KKpipi_cisi_BESIII}. The two uncertainties are statistical and systematic, respectively. The corresponding correlation matrix can be found in Ref.~\cite{cite:KKpipi_cisi_BESIII}.}
  \begin{center}
    \begin{tabular}{c*{2}{r@{\:$\pm$\:}c@{\:$\pm$\:}l}}
      \hline
      $i$ & \multicolumn{3}{c}{$c_i$}                                    & \multicolumn{3}{c}{$s_i$} \\
      \hline
      $1$ & $-0.22\phantom{0}$            & $0.08\phantom{0}$  & $0.01$  & $-0.47$ & $0.22$ & $0.04$ \\
      $2$ & $\phantom{-}0.79\phantom{0}$  & $0.04\phantom{0}$  & $0.01$  & $-0.17$ & $0.16$ & $0.04$ \\
      $3$ & $\phantom{-}0.862$            & $0.029$            & $0.008$ & $0.26$  & $0.14$ & $0.02$ \\
      $4$ & $-0.39\phantom{0}$            & $0.08\phantom{0}$  & $0.01$  & $0.52$  & $0.24$ & $0.04$ \\
      \hline
    \end{tabular}
  \end{center}
  \label{table:cisi_KKpipi}
\end{table}

\begin{table}[!tb]
  \caption{Measured values of $c_i$ and $s_i$ for the $\D\to\pip\pim\pip\pim$ decay that are used in this analysis, taken from Ref.~\cite{cite:4pi_cisi_BESIII}. The two uncertainties are statistical and systematic, respectively. The corresponding correlation matrix can be found in Ref.~\cite{cite:4pi_cisi_BESIII}.}
  \begin{center}
    \begin{tabular}{c*{2}{r@{\:$\pm$\:}c@{\:$\pm$\:}l}}
      \hline
      $i$ & \multicolumn{3}{c}{$c_i$}    & \multicolumn{3}{c}{$s_i$} \\
      \hline
      $1$ & $0.119$  & $0.091$ & $0.021$ & $-0.424$ & $0.210$ & $0.043$ \\
      $2$ & $0.738$  & $0.044$ & $0.017$ & $-0.390$ & $0.161$ & $0.058$ \\
      $3$ & $0.808$  & $0.027$ & $0.012$ & $-0.250$ & $0.124$ & $0.030$ \\
      $4$ & $0.423$  & $0.059$ & $0.017$ & $0.857$  & $0.186$ & $0.074$ \\
      $5$ & $-0.273$ & $0.094$ & $0.025$ & $-0.225$ & $0.252$ & $0.081$ \\
      \hline
    \end{tabular}
  \end{center}
  \label{table:cisi_4pi}
\end{table}

When represented on a two-dimensional plane $(c_i, s_i)$, the values are generally located far from the origin and close to the unit circle. This is an indication that the dilution of the strong-phase difference, when integrating over each bin, is minimal. Thus, the sensitivity to $\gamma$ is close to optimal. Conversely, if the strong-phase difference had shown significant variations within a phase-space bin, the values of $c_i$ and $s_i$, which are amplitude-averaged quantities, would approach zero when averaging the sinusoidal oscillations. The resulting effect on Eqs.~\eqref{equation:Bp_yield} and \eqref{equation:Bm_yield} would be a dilution of the interference terms, which reduces the sensitivity to $\gamma$.

\section{The LHCb detector and data set}
\label{section:The_LHCb_detector_and_data_set}
This analysis uses data collected by the \lhcb experiment in proton-proton ($pp$) collisions at centre-of-mass-energies of $\sqrt{s}= 7$, $8$ and $13\tev$, corresponding to integrated luminosities of $1$, $2$ and $6\invfb$, respectively.

The \lhcb detector~\cite{LHCb-DP-2008-001,LHCb-DP-2014-002} is a single-arm forward spectrometer covering the \mbox{pseudorapidity} range $2<\eta <5$, designed for the study of particles containing \bquark or \cquark quarks. The detector used to collect the data analysed in this paper includes a high-precision tracking system consisting of a silicon-strip vertex detector surrounding the $pp$ interaction region, a large-area silicon-strip detector located upstream of a dipole magnet with a bending power of about $4{\mathrm{\,T\,m}}$, and three stations of silicon-strip detectors and straw drift tubes placed downstream of the magnet. The tracking system provides a measurement of the momentum, \ptot, of charged particles with a relative uncertainty that varies from 0.5\% at low momentum to 1.0\% at 200\gevc. The minimum distance of a track to a primary $pp$ collision vertex (PV), the impact parameter (IP), is measured with a resolution of $(15+29/\pt)\mum$, where \pt is the component of the momentum transverse to the beam, in\,\gevc. Different types of charged hadrons are distinguished using information from two ring-imaging Cherenkov detectors. Photons, electrons and hadrons are identified by a calorimeter system consisting of scintillating-pad and preshower detectors, an electromagnetic and a hadronic calorimeter. Muons are identified by a system composed of alternating layers of iron and multiwire proportional chambers. The online event selection is performed by a trigger, which consists of a hardware stage, based on information from the calorimeter and muon systems, followed by a software stage, which applies a full event reconstruction.

Simulation is required to model the effects of the detector acceptance, and for the study of possible background processes. Furthermore, simulation samples are used as a training dataset in the multivariate selection described in Sect.~\ref{section:Selection_of_B_candidates}. In the simulation, $pp$ collisions are generated using \pythia~\cite{Sjostrand:2007gs,*Sjostrand:2006za} with a specific \lhcb configuration~\cite{LHCb-PROC-2010-056}. Decays of unstable particles are described by \evtgen~\cite{Lange:2001uf}, in which final-state radiation is generated using \photos~\cite{davidson2015photos}. The $\Dz\to\Kp\Km\pip\pim$ decay is simulated using the amplitude model from Ref.~\cite{LHCb-PAPER-2018-041}, while simulation of the $\Dz\to\pip\pim\pip\pim$ decay is not required. The interaction of the generated particles with the detector, and its response, are implemented using the \geant toolkit~\cite{Allison:2006ve, *Agostinelli:2002hh} as described in Ref.~\cite{LHCb-PROC-2011-006}. The underlying $pp$ interaction is reused multiple times, each with an independently generated signal decay~\cite{LHCb-DP-2018-004}.

\section{Selection of \texorpdfstring{\boldmath{$\Bpm$}}{B} candidates}
\label{section:Selection_of_B_candidates}
The selection of $\Bpm\to[\Kp\Km\pip\pim]_\D h^\pm$ and $\Bpm\to[\pip\pim\pip\pim]_\D h^\pm$ candidates is nearly identical to that described in Ref.~\cite{LHCb-PAPER-2022-037}. Five charged tracks are combined into a $\Bpm$ candidate, where it is required that four of these tracks, under the relevant pion or kaon mass hypotheses, have an invariant mass within $25\mevcc$ of the $\D$ meson mass~\cite{PDG2024}. This requirement, which corresponds to around two and a half times the resolution of the mass peak, is highly efficient at removing background processes that have either a missing or misidentified particle originating from the $\D$ meson. Additionally, there is a set of requirements to suppress other background decays that have peaking structures in the invariant mass of the $\Bpm$ candidate, and a multivariate selection to reduce the combinatorial background.

To suppress charmless background, which arises from $\Bpm$ meson decays where there is no intermediate charm meson, the distance between the $\D$ and $\Bpm$ decay vertices is required to be greater than twice its resolution. This criterion eliminates $95\%$ of charmless decays. Candidates where the opening angle between any pair of tracks from the $\D$-decay products is smaller than $0.03^\circ$ are discarded, as these are likely to correspond to a single charged particle that is duplicated in the reconstruction (cloned tracks).

Background from $\D$ decays where a $\pip\pim$ pair originates from a $\KS$ meson is suppressed by excluding events containing $\pip\pim$ pairs with an invariant mass consistent with a $\KS$ meson. For the $\D\to\Kp\Km\pip\pim$ mode, these are candidates inside the interval $[477, 507]\mevcc$~\cite{LHCb-PAPER-2022-037}, while the exclusion region for the $\D\to\pip\pim\pip\pim$ decay is $[480, 505]\mevcc$~\cite{cite:4pi_cisi_CLEOc}.

Separation of $\Bpm\to\D\Kpm$ and $\Bpm\to\D\pipm$ candidates is achieved by imposing mutually exclusive particle identification (PID) requirements on the companion $\Kpm$ or $\pipm$ meson from the $\Bpm\to\D h^\pm$ decay. Companion tracks that have associated activity in the muon detector are removed; this requirement reduces background from semileptonic $b$-hadron decays involving a muon that is misidentified as a kaon or pion. Background from semileptonic $b$-decays involving an electron is found to be negligible.

To distinguish between candidates where the subsequent $\D$ meson decays to a $\Kp\Km\pip\pim$ and a $\pip\pim\pip\pim$ final state, loose PID requirements are imposed between pions and kaons. Additionally, in the $\D\to\Kp\Km\pip\pim$ mode, tighter PID requirements are imposed on the kaon from the $\D$ candidate with opposite sign to the companion track. This selection requirement suppresses $\D\to\Kmp\pipm\pim\pip\piz$ background where the $\piz$ meson is not reconstructed and the kaon is misidentified. Furthermore, in the $\D\to\pip\pim\pip\pim$ channel, the pion PID requirements are tightened, with respect to Ref.~\cite{LHCb-PAPER-2022-037}, to remove combinatorial background from tracks not consistent with pions. This additional requirement is useful when analysing regions of phase space with small signal yields.

The majority of combinatorial background events are suppressed using a boosted decision tree~(BDT) classifier~\cite{Breiman,AdaBoost} implemented in the TMVA toolkit~\cite{Hocker:2007ht,*TMVA4}. Simulated signal $\Bpm\to[\Kp\Km\pip\pim]_\D h^\pm$ events are used as the signal training sample, while \Bpm candidates with an invariant mass in the range $[5800, 7000]\mevcc$ form the background training sample. The input variables of the BDT include the momenta and IPs of the $\Bpm$, $\D$ and companion-track candidates. The variables are described in detail in Ref.~\cite{LHCb-PAPER-2018-017}. The optimal working point of the BDT is chosen by minimising the expected statistical uncertainty on $\gamma$, based on studies using pseudoexperiments. While the BDT setup is identical to that described in Ref.~\cite{LHCb-PAPER-2022-037}, in the present analysis it is applied at the end of the selection procedure, and the BDT was retrained to adapt to the new selection procedure. In comparison, in Ref.~\cite{LHCb-PAPER-2022-037}, the BDT was applied before the rectangular selections that suppress charmless background, cloned tracks and $\KS$ decays.

To improve the resolution of the momenta of the $\D$-decay products and the invariant mass of the $\Bpm$ candidate, a kinematic fit is performed in which the $\D$-meson candidate is constrained to its known mass~\cite{PDG2024}, and the $\Bpm$ candidate is constrained to originate from its associated PV. This is defined as the PV with the smallest IP with respect to the $\Bpm$ candidate. After the kinematic fit, it is found that a negligible number of candidates are outside the kinematic boundary of the phase space, and these candidates are therefore removed from the selection.

\section{Invariant-mass fit}
\label{section:Invariant_mass_fit}
A set of unbinned, extended maximum-likelihood fits are performed simultaneously to the reconstructed invariant-mass spectra, $m(\D h^\pm)$, of the \mbox{$\Bpm\to[\Km\Kp\pip\pim]_\D h^\pm$} and \mbox{$\Bpm\to[\pim\pip\pip\pim]_\D h^\pm$} candidates in the range $[5080,5700]\mevcc$. The first fit, which is referred to as a global fit, is identical to that described in Ref.~\cite{LHCb-PAPER-2022-037}. It is performed on the selected $\Bpm\to\D\Kpm$ and $\Bpm\to\D\pipm$ candidates, integrated over all phase-space bins. The purpose of the global fit is to determine the parameters of the functions that describe the signal and background invariant-mass distributions. In subsequent fits, where these parameters are fixed, the \CP-violating behaviour is studied in phase-space bins and the resulting \CP-violating observables are extracted. The final maximum-likelihood fit, which is discussed in Sect.~\ref{section:Interpretation_in_terms_of_gamma_and_other_physics_parameters}, uses the results from this section to interpret the yields of each phase-space bin in terms of the physics parameters.

\begin{figure}[tb]
  \begin{center}
    \includegraphics[width=1.0\linewidth]{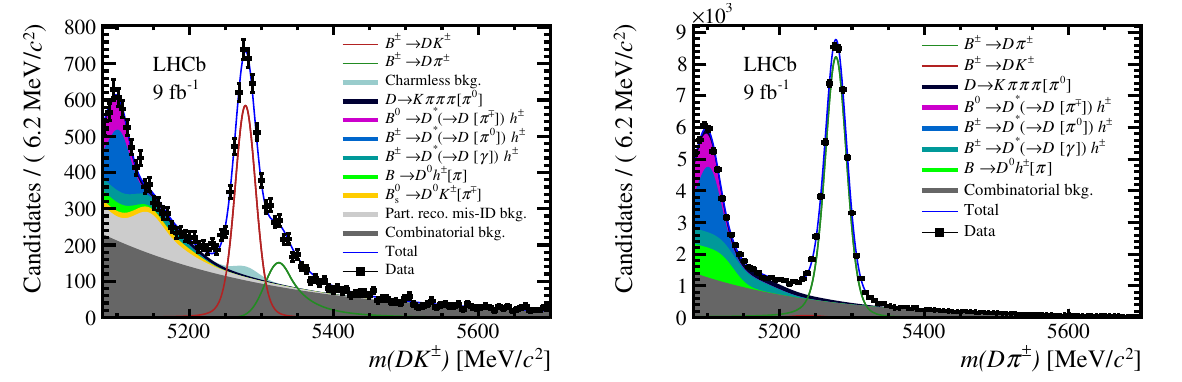}
  \end{center}
  \caption{Invariant-mass distributions for the (left) $\Bpm\to\D\Kpm$ and (right) $\Bpm\to\D\pipm$ selections, for the $\D\to\Kp\Km\pip\pim$ decay. The square brackets in the legend denote particles that are not reconstructed. With the exception of the (red) $\Bpm\to\D\Kpm$ and (green) $\Bpm\to\D\pipm$ components, the distributions are stacked.}
  \label{figure:Global_fit_KKpipi}
\end{figure}

\begin{figure}[tb]
  \begin{center}
    \includegraphics[width=1.0\linewidth]{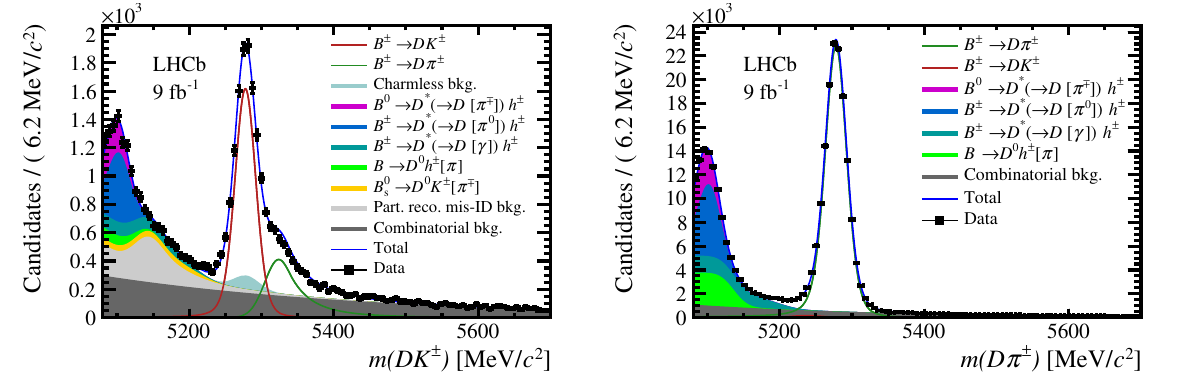}
  \end{center}
  \caption{Invariant-mass distributions for the (left) $\Bpm\to\D\Kpm$ and (right) $\Bpm\to\D\pipm$ selections, for the $\D\to\pip\pim\pip\pim$ decay. The square brackets in the legend denote particles that are not reconstructed. With the exception of the (red) $\Bpm\to\D\Kpm$ and (green) $\Bpm\to\D\pipm$ components, the distributions are stacked.}
  \label{figure:Global_fit_pipipipi}
\end{figure}

The invariant-mass distributions of $\Bpm\to[\Km\Kp\pip\pim]_\D h^\pm$ candidates are shown in Fig.~\ref{figure:Global_fit_KKpipi}, while those of $\Bpm\to[\pim\pip\pip\pim]_\D h^\pm$ are shown in Fig.~\ref{figure:Global_fit_pipipipi}. In the global fit, the yield of $\Bpm\to\D\pipm$ is varied separately for the two $\D$ decays, while the yield of $\Bpm\to\D\Kpm$ is parameterised as a ratio relative to the $\Bpm\to\D\pipm$ yield. This ratio is a common fit parameter for the two $\D$ decays, as are the parameters that describe the signal shape.

The peak at around $5280\mevcc$ corresponds to $\Bpm\to\D h^\pm$ candidates that are correctly reconstructed. The signal invariant-mass shape is parameterised as a sum of a Gaussian function and a modified Gaussian function, as detailed in Ref.~\cite{LHCb-PAPER-2022-037}. Similarly, at masses above the $\Bpm\to\D\Kpm$ peak there is a non-negligible contribution from $\Bpm\to\D\pipm$ decays where the companion is misidentified as a kaon. The treatment of this component is explained in Ref.~\cite{LHCb-PAPER-2022-037}, along with the analogous misidentification component of $\Bpm\to\D\Kpm$ candidates in the $\Bpm\to\D\pipm$ selection.

Candidates with masses below that of the signal peak are background from $\B$-meson decays where an additional pion or photon is not reconstructed. The model describing this partially reconstructed background, and its associated parameters, are taken from Ref.~\cite{LHCb-PAPER-2020-019}, with the exception of the contamination from $\Bs\to\Dzb\Km\pip$ and $\Bsb\to\Dz\Kp\pim$ decays with a pion that is not reconstructed. Instead, for this component, the total size of the background contamination is taken from the fit results of Ref.~\cite{LHCb-PAPER-2020-036}.

The contamination of charmless decays in the $\D\to\Kp\Km\pip\pim$ and $\pip\pim\pip\pim$ modes is also different from Ref.~\cite{LHCb-PAPER-2020-019}. In particular, the $\Bpm\to\D\Kpm$ sample has a significant charmless contribution. Although this type of background was not included as part of the baseline fit model in Ref.~\cite{LHCb-PAPER-2020-019}, it is necessary to do so for the four-body decays described here, due to the larger relative size of this background. Both the yield and invariant-mass shape of this contribution are fixed from studies of the lower (upper) sideband of the $\D\to\Kp\Km\pip\pim$ ($\pip\pim\pip\pim$) invariant mass. In these fits, the charmless contribution is modelled with a Gaussian function. With the exception of the combinatorial background, which is described by an exponential function, other sources of background are found to be negligible in the $\D$-sideband selection. Analogous studies show that there is no significant contamination from charmless decays in the $\Bpm\to\D\pipm$ selection.

Additionally, the $\D\to\Kp\Km\pip\pim$ mode is contaminated by $\D\to\Kmp\pipm\pim\pip\piz$ decays, where a charged pion is misidentified as a kaon and the neutral pion is not reconstructed. This background is present at mass values below the signal peak, but has a large tail which extends into the signal region. The shape of this background is fixed from simulation, while its yield is a free parameter. The ratio between this background and the signal is shared between the $\Bpm\to\D\Kpm$ and $\Bpm\to\D\pipm$ modes. In the absence of kaons in the final state, the $\D\to\pip\pim\pip\pim$ decay has no such partially reconstructed misidentified component.

The signal yields, obtained from the global fit, are given in Table~\ref{table:Global_fit_yields}. The yields are scaled from the full fit region to the signal region $m(\D h^\pm)\in[5249, 5309]\mevcc$ by integrating the relevant probability distributions. The uncertainties on the $\Bpm\to\D\Kpm$ yields are reduced due to the common ratio determined from both the $\D\to\Kp\Km\pip\pim$ and $\D\to\pip\pim\pip\pim$ decay modes.

\begin{table}[tb]
    \centering
    \caption{Yields of $\Bpm\to\D\Kpm$ and $\Bpm\to\D\pipm$ candidates, partially reconstructed background, $\D\to\Km\pip\pim\pip\piz$ background, combinatorial background and charmless background in the region $m(\D h^\pm)\in[5249, 5309]\mevcc$, where the charm meson decays via $\D\to\Kp\Km\pip\pim$ and $\D\to\pip\pim\pip\pim$. Charmless background in the $\Bpm\to\D\pipm$ decay modes is known to be negligible, and is therefore not modelled.}
    \label{table:Global_fit_yields}
    \begin{tabular}{ll*{2}{r@{\:$\pm$\:}l}} 
        \toprule
    &                              & \multicolumn{4}{c}{Reconstructed as:} \\
        $\D$ decay                   & Component                    & \multicolumn{2}{c}{$\Bpm\to\D\Kpm$} & \multicolumn{2}{c}{$\Bpm\to\D\pipm$} \\
        \midrule
        $\D\to~\Kp\Km\pip\pim$       & $\Bpm\to\D\Kpm$              & $3280$   & $41$     & $154$    & $2$      \\
    & $\Bpm\to\D\pipm$             & $258$    & $1$      & $47610$  & $230$    \\
    & Partially reconstructed bkg.  & $93$     & $1$      & $29$     & $1$      \\
    & $\D\to~\Km\pip\pim\pip\piz$  & $47$     & $13$     & $620$    & $170$    \\
    & Combinatorial bkg.            & $840$    & $24$     & $3540$   & $190$    \\
    & Charmless bkg.                & \multicolumn{2}{c}{$155$ (fixed)} & \multicolumn{2}{c}{Not modelled} \\
        \midrule
        $\D\to~\pip\pim\pip\pim$     & $\Bpm\to\D\Kpm$              & $9170$   & $110$    & $409$    & $5$      \\
   & $\Bpm\to\D\pipm$             & $708$    & $2$      & $132250$ & $390$    \\
   & Partially reconstructed bkg.  & $269$    & $2$      & $84$     & $3$      \\
   & Combinatorial bkg.            & $1594$   & $27$     & $4377$   & $82$     \\
   & Charmless bkg.                & \multicolumn{2}{c}{$629$ (fixed)} & \multicolumn{2}{c}{Not modelled} \\
        \bottomrule
    \end{tabular}
\end{table}

After the global invariant-mass fit, the $\Bpm\to\D\Kpm$ and $\Bpm\to\D\pipm$ candidates are split by charge and sorted into bins of phase space. The lower fit boundary is increased to $5150\mevcc$ to remove most of the partially reconstructed background. The shape parameters of the different background components are fixed from the global fit. The small variations in these background distributions between phase-space bins is considered as a systematic uncertainty, as discussed in Sect.~\ref{section:Systematic_uncertainties}.

In each bin, the yield of combinatorial background is a free parameter. Furthermore, the relative sizes of the partially reconstructed background components are fixed, such that only the total yield of partially reconstructed background in each bin is allowed to vary independently. The exception is the contamination from $\Bs\to\Dzb\Km\pip$ (and its charge conjugate), in which the charm meson has the opposite flavour to the signal decay. Consequently, for $\Bm$ ($\Bp$) candidates, the fractional bin yield of the $\Bs$ (\Bsb) background is set equal to $F_{-i}$ ($F_i$). In simulation, the $\D \to\Kmp\pipm\pim\pip\piz$ decays are uniformly distributed in the $\D \to\Kp\Km\pip\pim$ phase space. Therefore, the distribution of $\D \to\Kmp\pipm\pim\pip\piz$ decays between phase-space bins is assumed to be proportional to the bin volume. The distribution of the charmless background between phase-space bins is determined from the $\D$-mass sidebands.

\begin{figure}[tb]
    \centering
    \begin{subfigure}{0.5\textwidth}
        \includegraphics[width=1\textwidth]{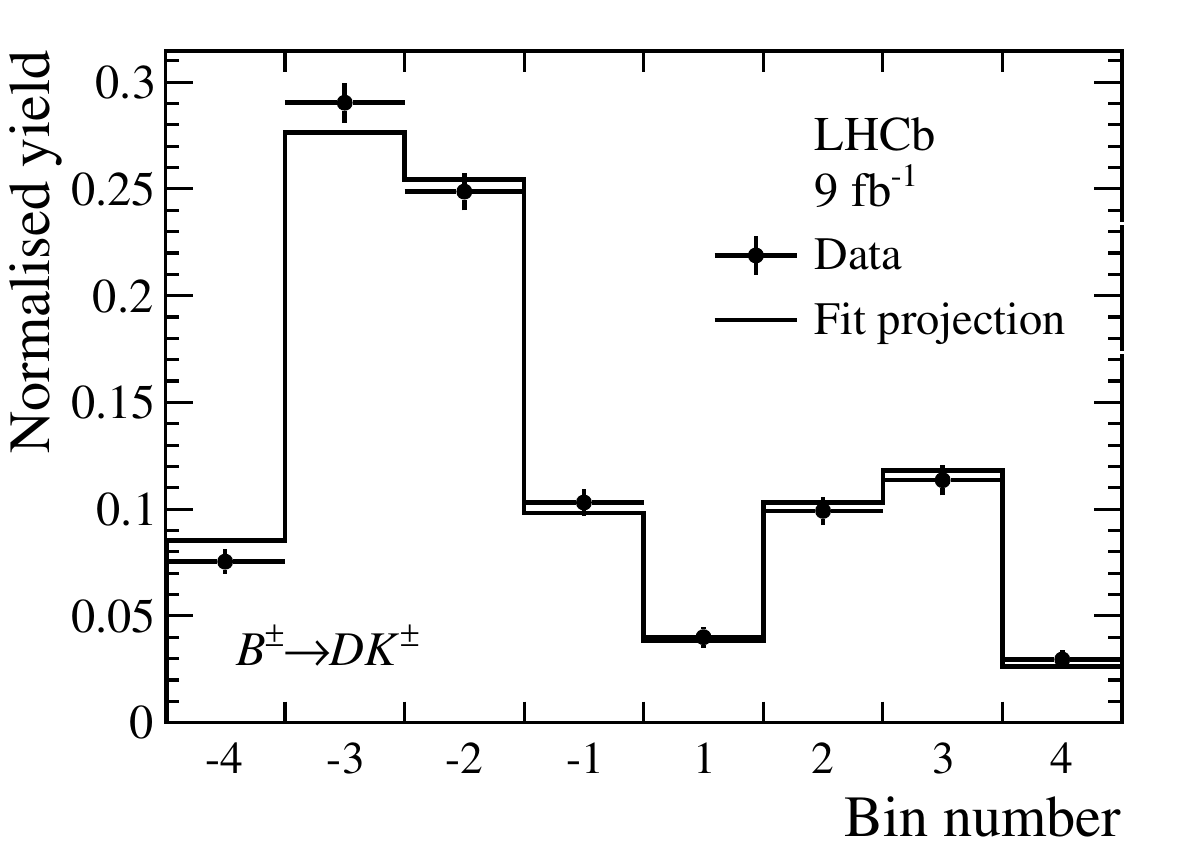}
    \end{subfigure}%
    \begin{subfigure}{0.5\textwidth}
        \includegraphics[width=1\textwidth]{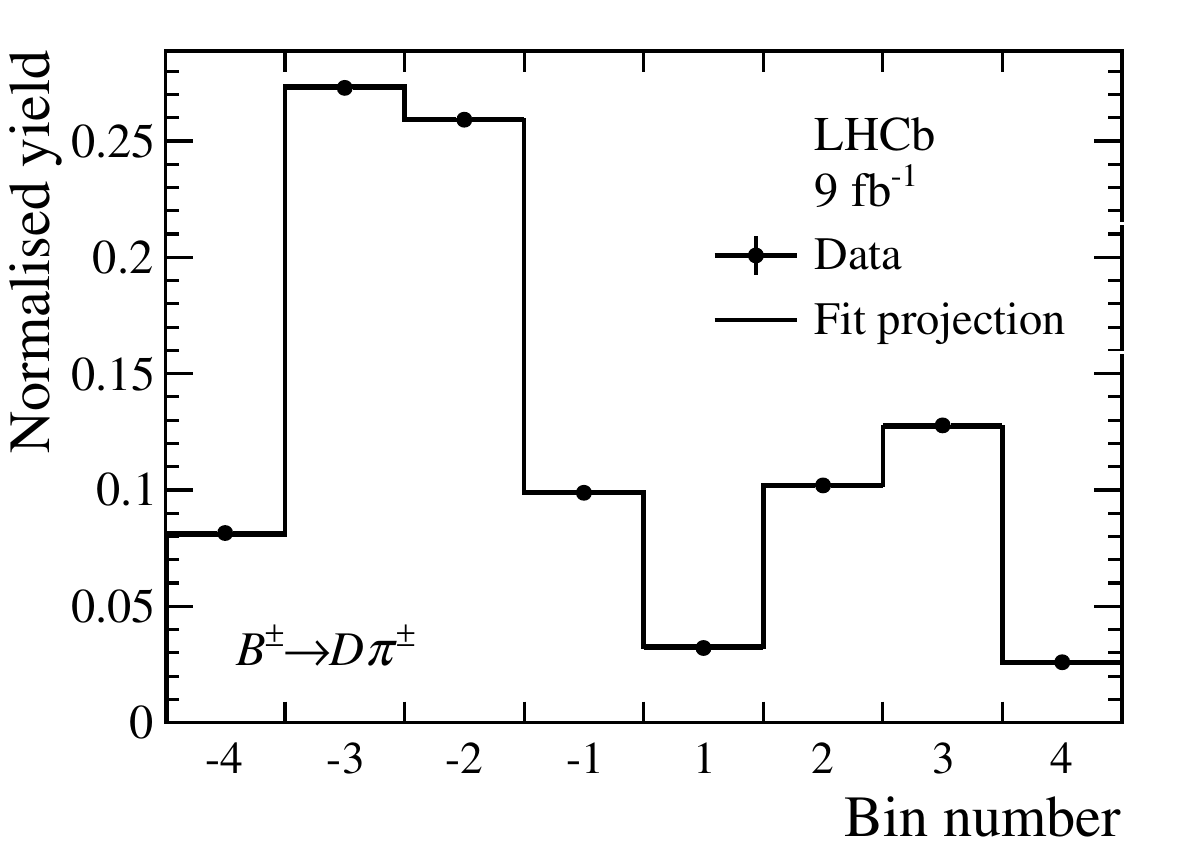}
    \end{subfigure}
    \caption{Signal yields for each phase-space bin, normalised to unity, for the (left) $\Bpm\to\D\Kpm$ and (right) $\Bpm\to\D\pipm$ candidates, with $\D\to\Kp\Km\pip\pim$. The fit projection is also shown.}
    \label{figure:Total_bin_yields_KKpipi}
\end{figure}

\begin{figure}[tb]
    \centering
    \begin{subfigure}{0.5\textwidth}
        \includegraphics[width=1\textwidth]{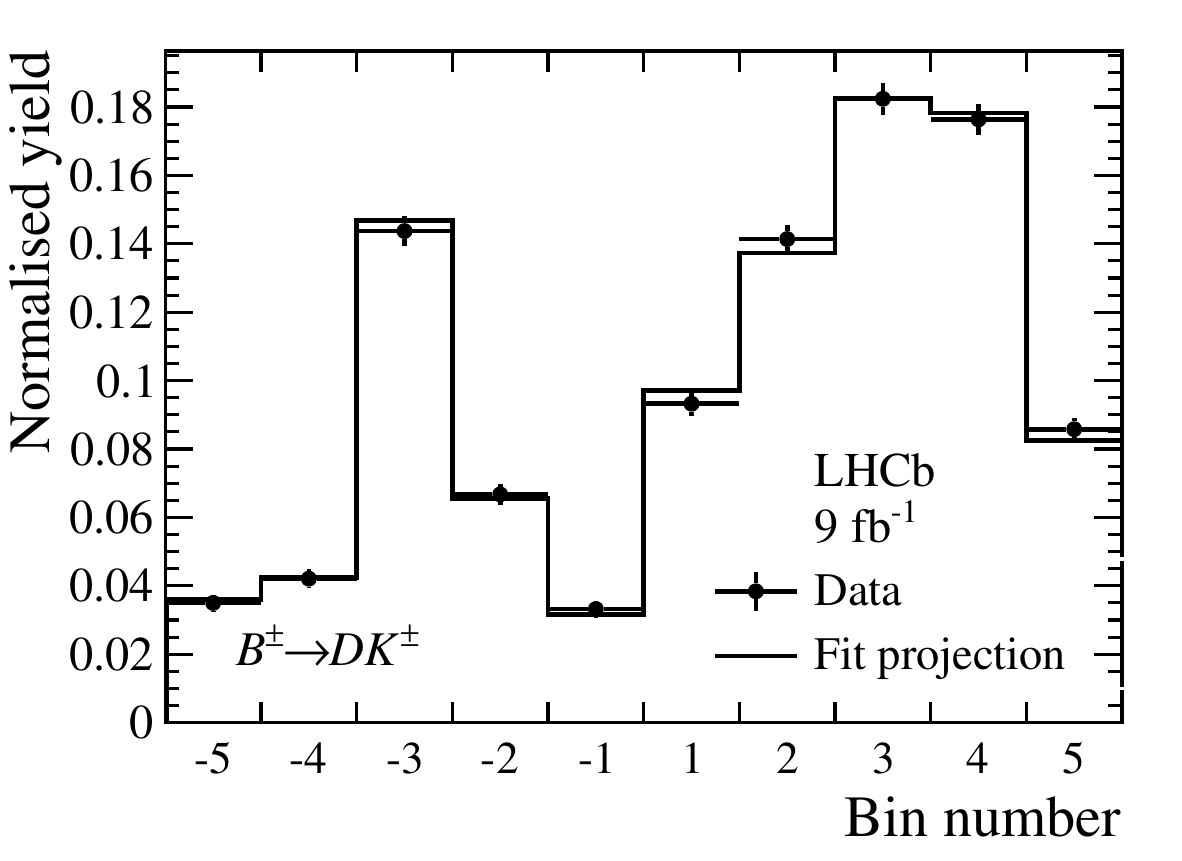}
    \end{subfigure}%
    \begin{subfigure}{0.5\textwidth}
        \includegraphics[width=1\textwidth]{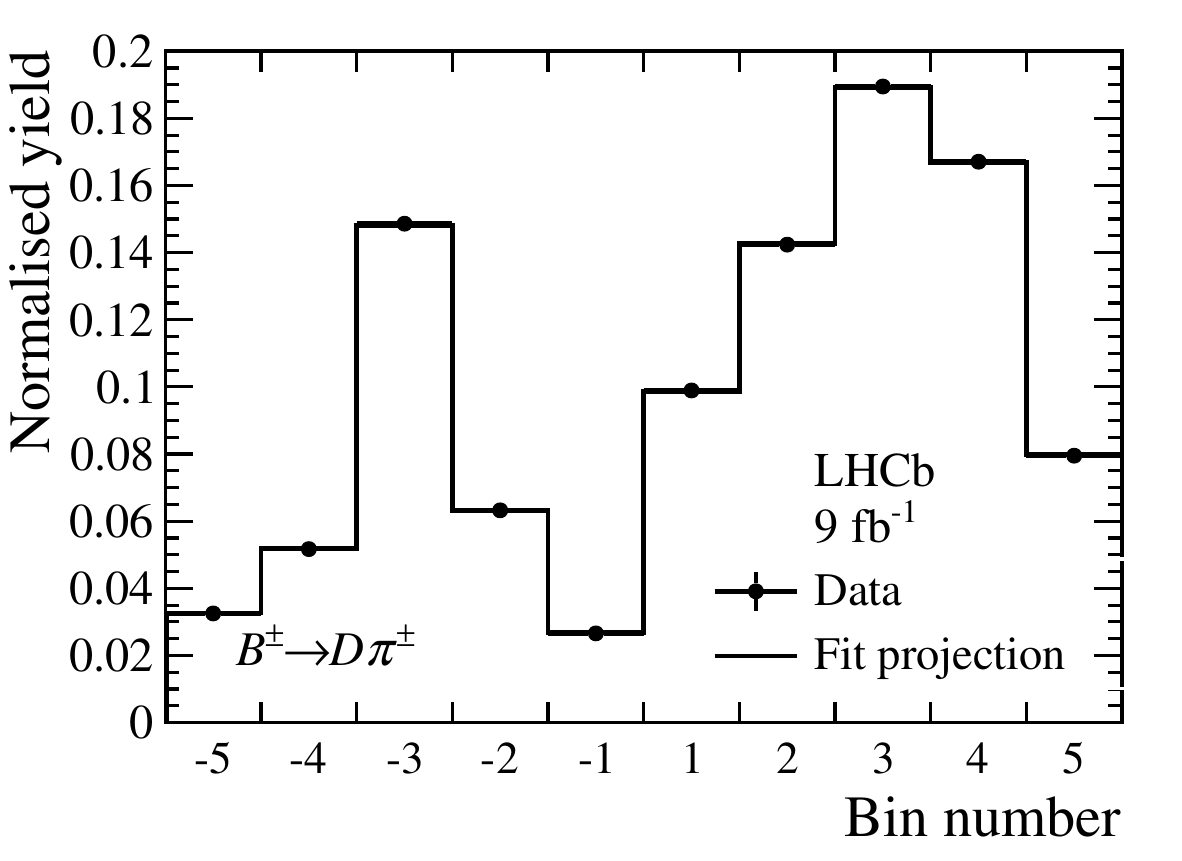}
    \end{subfigure}
    \caption{Signal yields for each phase-space bin, normalised to unity, for the (left) $\Bpm\to\D\Kpm$ and (right) $\Bpm\to\D\pipm$ candidates, with $\D\to\pip\pim\pip\pim$. The fit projection is also shown.}
    \label{figure:Total_bin_yields_pipipipi}
\end{figure}

There are two strategies for the treatment of the signal yields in each bin. The first strategy is to consider each signal yield as a free parameter, which serves as a useful consistency check. In Figs.~\ref{figure:Total_bin_yields_KKpipi} and \ref{figure:Total_bin_yields_pipipipi}, the sum of $\Bp$ and $\Bm$ yields, with the histogram normalised to unity, are shown as data points for each phase-space bin.

\begin{figure}[tb]
    \centering
    \begin{subfigure}{0.5\textwidth}
        \includegraphics[width=1\textwidth]{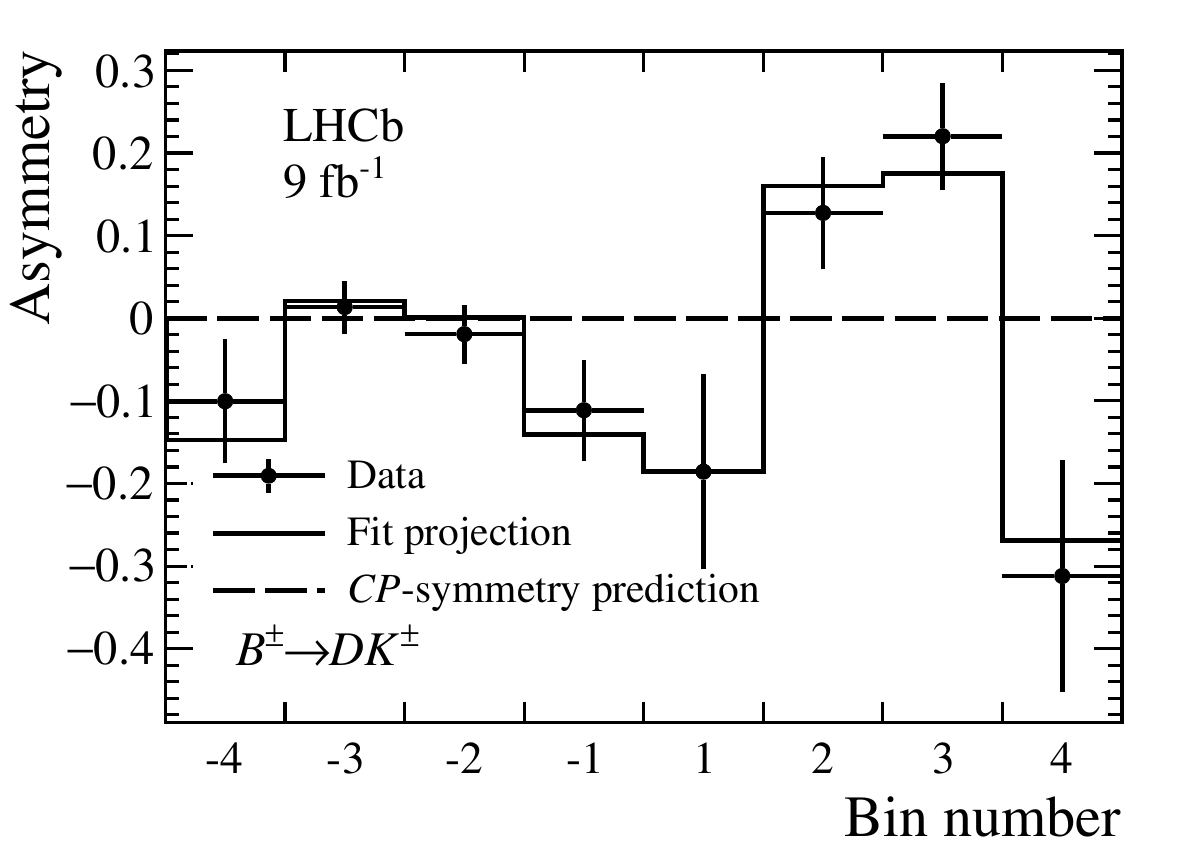}
    \end{subfigure}%
    \begin{subfigure}{0.5\textwidth}
        \includegraphics[width=1\textwidth]{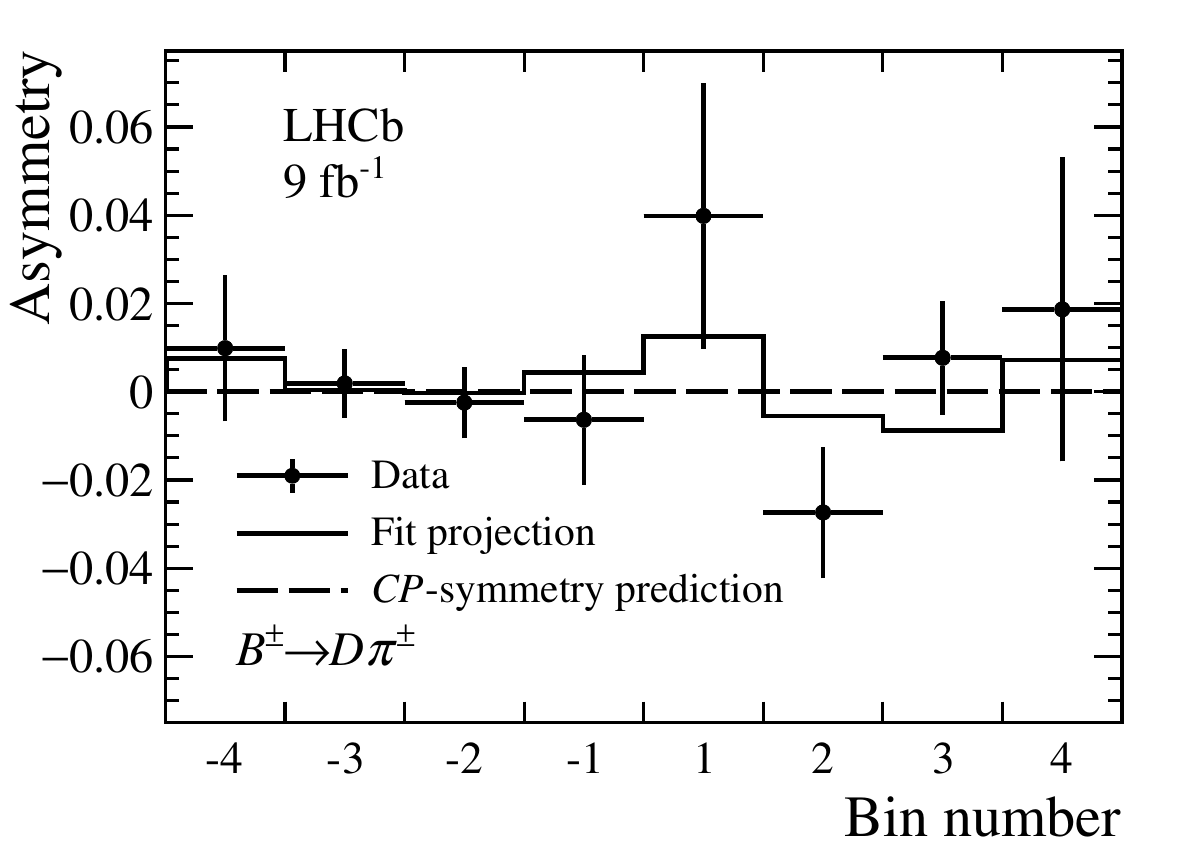}
    \end{subfigure}
    \caption{Fractional bin asymmetries for the (left) $\Bpm\to\D\Kpm$ and (right) $\Bpm\to\D\pipm$ decays, with $\D\to\Kp\Km\pip\pim$. The dashed line shows the prediction in the absence of \CP violation, and the fit projection is shown as a solid line.}
    \label{figure:Bin_asymmetries_KKpipi}
\end{figure}

\begin{figure}[tb]
    \centering
    \begin{subfigure}{0.5\textwidth}
        \includegraphics[width=1\textwidth]{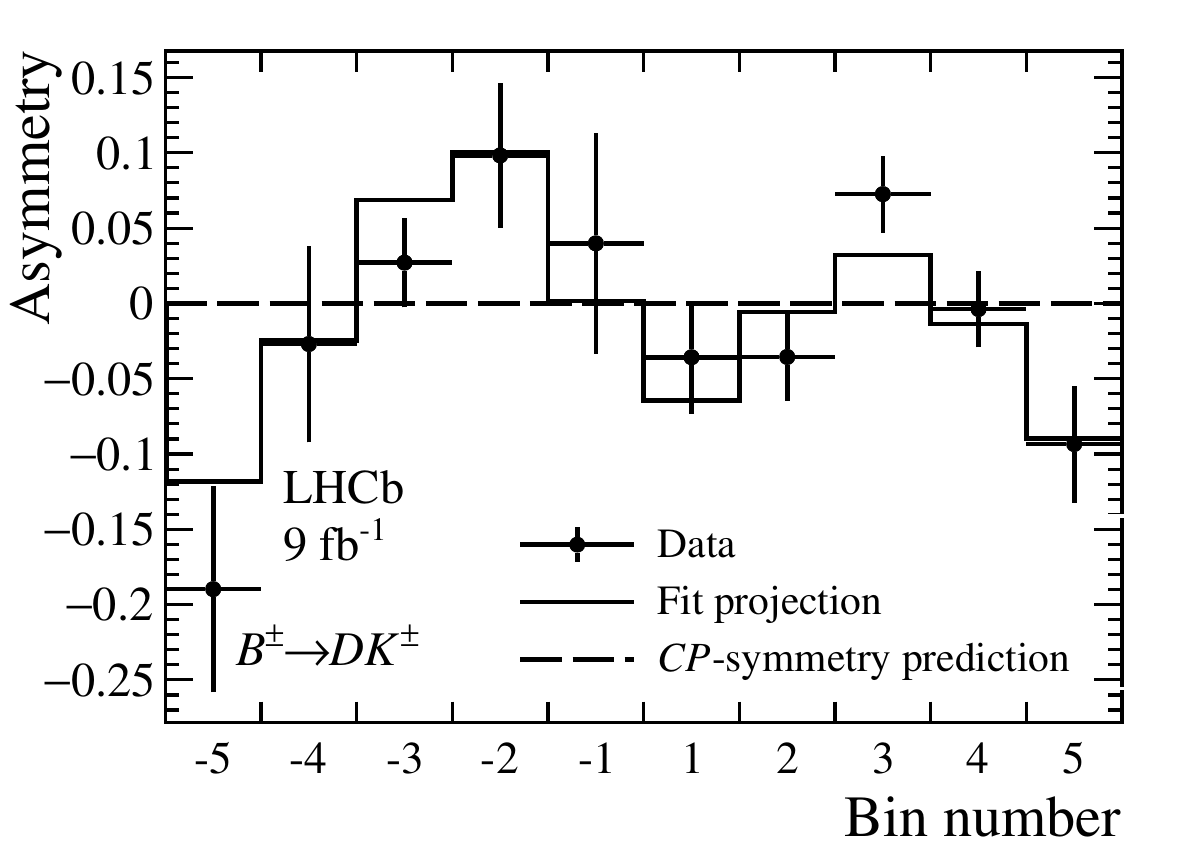}
    \end{subfigure}%
    \begin{subfigure}{0.5\textwidth}
        \includegraphics[width=1\textwidth]{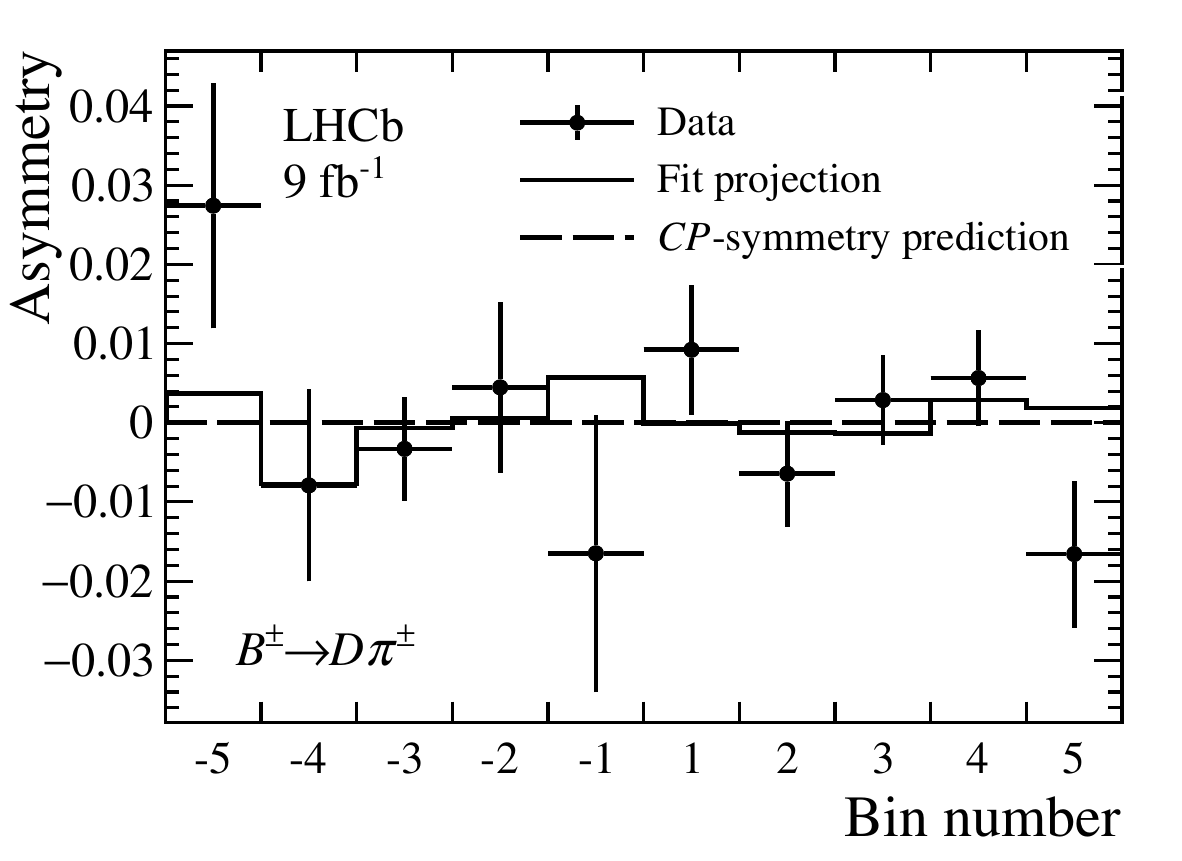}
    \end{subfigure}
    \caption{Fractional bin asymmetries for the (left) $\Bpm\to\D\Kpm$ and (right) $\Bpm\to\D\pipm$ decays, with $\D\to\pip\pim\pip\pim$. The dashed line shows the prediction in the absence of \CP violation, and the fit projection is shown as a solid line.}
    \label{figure:Bin_asymmetries_pipipipi}
\end{figure}

From the bin yields, it is possible to study directly the \CP-violation effects independent of the external strong-phase inputs by considering the per-bin asymmetry, which in bin $+i$ is defined as $(N_{+i}^- - N_{-i}^+)/(N_{+i}^- + N_{-i}^+)$. The bin yields are normalised separately for $\Bp$ and $\Bm$, so that only variations between different bins are present. The results are shown in Figs.~\ref{figure:Bin_asymmetries_KKpipi} and \ref{figure:Bin_asymmetries_pipipipi}, where nonzero asymmetries are seen for the $\Bpm\to\D\Kpm$ mode on the left. The nontrivial variation of bin asymmetries between different bins is driven by the different values of $c_i$ and $s_i$, which is caused by the variation in strong-phase difference across phase space. The $\Bpm\to\D\pipm$ mode, shown on the right, exhibits much smaller bin asymmetries due to the smaller value of $r_\B^{\D\pion}$, and the absence of observed \CP asymmetries in this mode indicates that no significant fake sources of \CP asymmetry are present.

The second strategy is to parameterise the signal yields in terms of the \CP-violating observables $x_{\pm}^{\D\kaon}$, $y_{\pm}^{\D\kaon}$, $x_{\xi}^{\D\pion}$ and $y_{\xi}^{\D\pion}$ using Eqs.~\eqref{equation:Bp_yield} and \eqref{equation:Bm_yield}. The observables are treated as free fit parameters that contain all information about the \CP-violating effects in the $\Bpm\to\D h^\pm$ decays. The solid line in Figs.~\ref{figure:Total_bin_yields_KKpipi}--\ref{figure:Bin_asymmetries_pipipipi} shows the fit result projected back onto the yields using Eqs.~\eqref{equation:Bp_yield} and \eqref{equation:Bm_yield}, with the fitted \CP-violating observables used as input. A good agreement between the measured yields and the fit projections is seen, which indicates that the fit model describes these decays well.

The motivation for this parameterisation is that the parameters $r_\B^{\D\kaon}$ and $r_\B^{\D\pion}$ are required to be non-negative, resulting in inaccurate uncertainty coverage near $r_\B^{\D h} = 0$. On the contrary, the \CP-violating observables have no such restriction, so these observables are generally well described by a multidimensional Gaussian distribution. Therefore, the subsequent interpretation in terms of the underlying physics parameters $\gamma$, $\delta_\B^{\D\kaon}$, $r_\B^{\D\kaon}$, $\delta_\B^{\D\pion}$ and $r_\B^{\D\pion}$ becomes more statistically robust. Furthermore, it makes the combination with other measurements sensitive to $\gamma$ more straightforward~\cite{LHCb-CONF-2024-004}.

Two key ingredients in the parameterisation of the signal yields in terms of the \CP-violating observables, or the physics parameters, are the $F_i$ parameters and the external inputs $c_i$ and $s_i$. The fractional bin yields $F_i$ are treated as free parameters, using the recursive fractions defined in Sect.~\ref{section:Analysis_strategy}. The advantages of this method have already been demonstrated in Refs.~\cite{LHCb-PAPER-2020-019} and \cite{LHCb-PAPER-2022-037}. The treatment of $c_i$ and $s_i$, on the other hand, is new. In previous measurements using the same formalism, such as Ref.~\cite{LHCb-PAPER-2020-019}, these external inputs were fixed. The uncertainties of $c_i$ and $s_i$ were then propagated separately and considered as a source of systematic uncertainty.

In this analysis, it is found that the systematic uncertainty on $\gamma$ that arises from the uncertainties on $s_i$ is of similar size to the statistical uncertainty. Hence, to account accurately for the effect of these external inputs on the central value and uncertainty of $\gamma$, the $c_i$ and $s_i$ parameters are constrained, with Gaussian priors, to their model-independent measurements from Refs.~\cite{cite:KKpipi_cisi_BESIII} and \cite{cite:4pi_cisi_BESIII}. Furthermore, since the large uncertainties of $s_i$ may induce non-Gaussian behaviour in the distribution of the \CP-violating observables, the interpretation is performed using the full log-likelihood function of these observables, which also includes other nuisance parameters such as the yield of combinatorial and low-mass background in each bin. This approach is the focus of Sect.~\ref{section:Interpretation_in_terms_of_gamma_and_other_physics_parameters}.

\section{Systematic uncertainties}
\label{section:Systematic_uncertainties}
There are several sources of systematic uncertainties associated with the measured value of $\gamma$ and the other physics parameters. The contribution of each systematic uncertainty, and their correlations, is evaluated for each of the \CP-violating observables, and then propagated to the physics parameters. Table~\ref{table:Systematic_uncertainties_KKpipi} lists all the sources of systematic uncertainty and the size of each systematic for each \CP-violating observable in the analysis of the $\D\to\Kp\Km\pip\pim$ mode.

The procedure for evaluating each contribution is identical to that described in Ref.~\cite{LHCb-PAPER-2022-037}, and the values of the assignments in Table~\ref{table:Systematic_uncertainties_KKpipi} are therefore identical to those of Ref.~\cite{LHCb-PAPER-2022-037}, with two exceptions. First, the component that describes the model dependence of $c_i$ and $s_i$ has been removed, as these parameters are now constrained to their model-independent values, using the full covariance matrix. Therefore, the uncertainties of $c_i$ and $s_i$ are already accounted for in the statistical uncertainties of the physics parameters. Second, the uncertainties assigned for fit bias have been re-evaluated, as these biases are different due to the influence of the $c_i$ and $s_i$ parameters.

The evaluation of the systematic uncertainties in the $\D\to\pip\pim\pip\pim$ channel is identical to that of the $\D\to\Kp\Km\pip\pim$ channel, but since the $\D\to\kaon^\mp\pion^\pm\pion^+\pion^-$, $\D\to\kaon^\mp\pion^\pm\pion^+\pion^-\piz$ and semileptonic charm-decay backgrounds are not present in the selected sample for this channel, no corresponding uncertainty has been assigned. These numbers are shown in Table~\ref{table:Systematic_uncertainties_pipipipi}. Finally, the systematic uncertainties for the simultaneous fit of these two charm decays are shown in Table~\ref{table:Systematic_uncertainties_hhpipi}.

\begin{table}[tb]
    \centering
    \caption{Systematic uncertainties on the results of the binned analysis in the $\D\to\Kp\Km\pip\pim$ channel.}
    \label{table:Systematic_uncertainties_KKpipi}
    \begin{tabular}{lcccccc}
        \toprule
        & \multicolumn{6}{c}{Uncertainty ($\times 10^2$)} \\
        \midrule
        Source & $x_-^{\D\kaon}$ & $y_-^{\D\kaon}$ & $x_+^{\D\kaon}$ & $y_+^{\D\kaon}$ & $x_\xi^{\D\pion}$ & $y_\xi^{\D\pion}$ \\
        \midrule
        Mass shape                                              & $0.01$ & $0.01$ & $0.02$ & $0.04$ & $0.02$ & $0.03$ \\
        Bin-dependent mass shape                                & $0.31$ & $0.43$ & $0.36$ & $0.10$ & $0.40$ & $0.01$ \\
        PID efficiency                                          & $0.01$ & $0.01$ & $0.02$ & $0.04$ & $0.02$ & $0.03$ \\
        Low-mass background model                               & $0.01$ & $0.01$ & $0.01$ & $0.01$ & $0.03$ & $0.01$ \\
        Charmless background                                    & $0.11$ & $0.17$ & $0.09$ & $0.14$ & $0.01$ & $0.01$ \\
        \CP violation in low-mass background                    & $0.05$ & $0.22$ & $0.07$ & $0.14$ & $0.15$ & $0.13$ \\
        Semileptonic $\bquark$-hadron decays                    & $0.07$ & $0.08$ & $0.03$ & $0.07$ & $0.17$ & $0.56$ \\
        Semileptonic $\cquark$-hadron decays                    & $0.02$ & $0.01$ & $0.04$ & $0.02$ & $0.15$ & $0.29$ \\
        $\D\to\kaon^\mp\pion^\pm\pion^+\pion^-$ background      & $0.02$ & $0.09$ & $0.01$ & $0.06$ & $0.25$ & $0.61$ \\
        $\Lb \to p D \pi^-$ background                          & $0.16$ & $0.04$ & $0.13$ & $0.04$ & $0.01$ & $0.12$ \\
        $\D\to\kaon^\mp\pion^\pm\pion^+\pion^-\piz$ background  & $0.09$ & $0.11$ & $0.44$ & $0.08$ & $0.12$ & $0.46$ \\
        Fit bias                                                & $0.13$ & $0.13$ & $0.10$ & $0.16$ & $0.10$ & $0.12$ \\
        \midrule
        Total \lhcb systematic                                  & $0.41$ & $0.56$ & $0.61$ & $0.31$ & $0.57$ & $1.02$ \\
        \bottomrule
    \end{tabular}
\end{table}

\begin{table}[tb]
    \centering
    \caption{Systematic uncertainties on the results of the binned analysis in the $\D\to\pip\pim\pip\pim$ channel.}
    \label{table:Systematic_uncertainties_pipipipi}
    \begin{tabular}{lcccccc}
        \toprule
        & \multicolumn{6}{c}{Uncertainty ($\times 10^2$)} \\
        \midrule
        Source & $x_-^{\D\kaon}$ & $y_-^{\D\kaon}$ & $x_+^{\D\kaon}$ & $y_+^{\D\kaon}$ & $x_\xi^{\D\pion}$ & $y_\xi^{\D\pion}$ \\
        \midrule
        Mass shape                                              & $0.03$ & $0.02$ & $0.03$ & $0.03$ & $0.06$ & $0.04$ \\
        Bin-dependent mass shape                                & $0.03$ & $0.05$ & $0.33$ & $0.00$ & $0.29$ & $0.23$ \\
        PID efficiency                                          & $0.03$ & $0.02$ & $0.03$ & $0.03$ & $0.06$ & $0.04$ \\
        Low-mass background model                               & $0.01$ & $0.01$ & $0.01$ & $0.01$ & $0.05$ & $0.04$ \\
        Charmless background                                    & $0.12$ & $0.09$ & $0.12$ & $0.10$ & $0.05$ & $0.04$ \\
        \CP violation in low-mass background                    & $0.11$ & $0.02$ & $0.16$ & $0.08$ & $0.12$ & $0.14$ \\
        Semileptonic $\bquark$-hadron decays                    & $0.13$ & $0.02$ & $0.14$ & $0.07$ & $0.04$ & $0.02$ \\
        $\Lb \to p D \pi^-$ background                          & $0.23$ & $0.08$ & $0.01$ & $0.04$ & $0.06$ & $0.09$ \\
        Fit bias                                                & $0.07$ & $0.06$ & $0.06$ & $0.07$ & $0.02$ & $0.10$ \\
        \midrule
        Total \lhcb systematic                                  & $0.32$ & $0.15$ & $0.42$ & $0.17$ & $0.34$ & $0.32$ \\
        \bottomrule
    \end{tabular}
\end{table}

\begin{table}[tb]
    \centering
    \caption{Systematic uncertainties on the results of the simultaneous binned analysis of the $\D\to\Kp\Km\pip\pim$ and $\D\to\pip\pim\pip\pim$ channels.}
    \label{table:Systematic_uncertainties_hhpipi}
    \begin{tabular}{lcccccc}
        \toprule
        & \multicolumn{6}{c}{Uncertainty ($\times 10^2$)} \\
        \midrule
        Source & $x_-^{\D\kaon}$ & $y_-^{\D\kaon}$ & $x_+^{\D\kaon}$ & $y_+^{\D\kaon}$ & $x_\xi^{\D\pion}$ & $y_\xi^{\D\pion}$ \\
        \midrule
        Mass shape                                              & $0.02$ & $0.02$ & $0.03$ & $0.02$ & $0.01$ & $0.01$ \\
        Bin-dependent mass shape                                & $0.10$ & $0.26$ & $0.06$ & $0.01$ & $0.17$ & $0.09$ \\
        PID efficiency                                          & $0.02$ & $0.02$ & $0.03$ & $0.02$ & $0.01$ & $0.01$ \\
        Low-mass background model                               & $0.01$ & $0.01$ & $0.02$ & $0.01$ & $0.02$ & $0.01$ \\
        Charmless background                                    & $0.09$ & $0.08$ & $0.08$ & $0.09$ & $0.01$ & $0.01$ \\
        \CP violation in low-mass background                    & $0.14$ & $0.04$ & $0.03$ & $0.01$ & $0.06$ & $0.07$ \\
        Semileptonic $\bquark$-hadron decays                    & $0.10$ & $0.00$ & $0.02$ & $0.03$ & $0.12$ & $0.15$ \\
        Semileptonic $\cquark$-hadron decays                    & $0.09$ & $0.11$ & $0.01$ & $0.00$ & $0.28$ & $0.02$ \\
        $\D\to\kaon^\mp\pion^\pm\pion^+\pion^-$ background      & $0.12$ & $0.04$ & $0.03$ & $0.03$ & $0.28$ & $0.13$ \\
        $\Lb \to p D \pi^-$ background                          & $0.16$ & $0.01$ & $0.14$ & $0.03$ & $0.19$ & $0.06$ \\
        $\D\to\kaon^\mp\pion^\pm\pion^+\pion^-\piz$ background  & $0.06$ & $0.05$ & $0.18$ & $0.03$ & $0.10$ & $0.12$ \\
        Fit bias                                                & $0.07$ & $0.06$ & $0.05$ & $0.07$ & $0.09$ & $0.08$ \\
        \midrule
        Total \lhcb systematic                                  & $0.33$ & $0.31$ & $0.26$ & $0.13$ & $0.51$ & $0.28$ \\
        \bottomrule
    \end{tabular}
\end{table}

\section{Interpretation in terms of \texorpdfstring{$\gamma$}{gamma} and other physics parameters}
\label{section:Interpretation_in_terms_of_gamma_and_other_physics_parameters}
The numerical result of $\gamma$ is obtained in the final part of the unbinned maximum-likelihood fit. The binned signal yields of the $\Bpm\to[\Kp\Km\pip\pim]_\D h^\pm$ and $\Bpm\to[\pip\pim\pip\pim]_\D h^\pm$ decays, which are parameterised in terms of the \CP-violating observables, are interpreted in terms of $\gamma$ and the other physics parameters. As discussed at the end of Sect.~\ref{section:Invariant_mass_fit}, this analysis must make use of the full unbinned log-likelihood function, denoted by $\mathcal{L}(X_i, Z_k)$, where $X_i$ are the \CP-violating observables $x_\pm^{\D\kaon}$, $y_\pm^{\D\kaon}$, $x_\xi^{\D\pion}$ and $y_\xi^{\D\pion}$, and $Z_k$ are all the nuisance parameters, which are different for the two charm decays.

In particular, in the $\Bpm\to[\Kp\Km\pip\pim]_\D h^\pm$ decay, there are two $\Bpm$ decays, two charges and $2\times4$ phase-space bins, making a total of $32$ categories. Although the invariant-mass shapes and relative sizes of background components in each bin are fixed from the global fit, the total yield of the low-mass background and the combinatorial background in each bin are free parameters, making a total of $64$ nuisance parameters that describe the background yields. Furthermore, there are seven independent $F_i$ parameters, eight $c_i$ and $s_i$ parameters and four normalisation parameters $h_{\Bpm}^{\D h}$. 
In total, there are $83$ nuisance parameters describing the $\D\to\Kp\Km\pip\pim$ mode, and  $103$ nuisance parameters in the $\D\to\pip\pim\pip\pim$ channel.

In the interpretation, the \CP-violating observables $X_i$ are transformed into a set of new parameters $Y_j$, which are the physics parameters $\gamma$, $\delta_\B^{\D\kaon}$, $r_\B^{\D\kaon}$, $\delta_\B^{\D\pion}$ and $r_\B^{\D\pion}$. These are allowed to vary freely, and the transformation is given by the relation
\begin{equation}
    X_i = F_i(Y_j) + S_i,
    \label{equation:interpretation}
\end{equation}
where $F_i$ denotes the transformations given by Eqs.~\eqref{equation:Cartesian_parametersiation} and \eqref{equation:xi}. The term $S_i$ in Eq.~\eqref{equation:interpretation} is a multidimensional Gaussian distribution with zero mean and a covariance matrix obtained from Sect.~\ref{section:Systematic_uncertainties}.

The transformation in Eq.~\eqref{equation:interpretation} reduces the number of parameters from six to five, which effectively introduces the constraint $(x_+^{\D\kaon})^2 + (y_+^{\D\kaon})^2 = (x_-^{\D\kaon})^2 + (y_-^{\D\kaon})^2$. The two $\D$-decay channels may be fitted separately to determine these physics parameters, along with their nuisance parameters. In addition, a simultaneous fit of both $\D$ decays is also performed, where the nuisance parameters are kept separate, but the physics parameters are shared. The presence of the $S_i$ term accounts for systematic uncertainties directly in the interpretation, including correlation effects. In the absence of correlations associated with the systematic uncertainties, the addition of $S_i$ to the \CP-violating observables is equivalent to combining statistical and systematic uncertainties in quadrature.

Finally, the transformed likelihood function $\mathcal{L}(Y_j, Z_k)$ is maximised to obtain the measured physics parameters. While it is possible to obtain symmetric uncertainties on each parameter from the second derivative at the extremum, it is much more informative to assess the two-dimensional contours. Confidence regions are obtained by scanning the profiled log-likelihood $\ln(\mathcal{L})$, which is defined to be zero at its extremum location. From Wilks' theorem~\cite{Wilks:1938dza}, the quantity $\Delta\chi^2 \equiv -2\ln(\mathcal{L})$ follows a $\chi^2$ distribution. Hence, in two dimensions, a $68.3\%$ confidence region corresponds to $\Delta\chi^2 = 2.30$. Similarly, the $95.4\%$ confidence region is obtained by determining $\Delta\chi^2 = 6.18$.

\begin{figure}[tb]
    \centering
    \begin{subfigure}{0.5\textwidth}
        \includegraphics[width=1\textwidth]{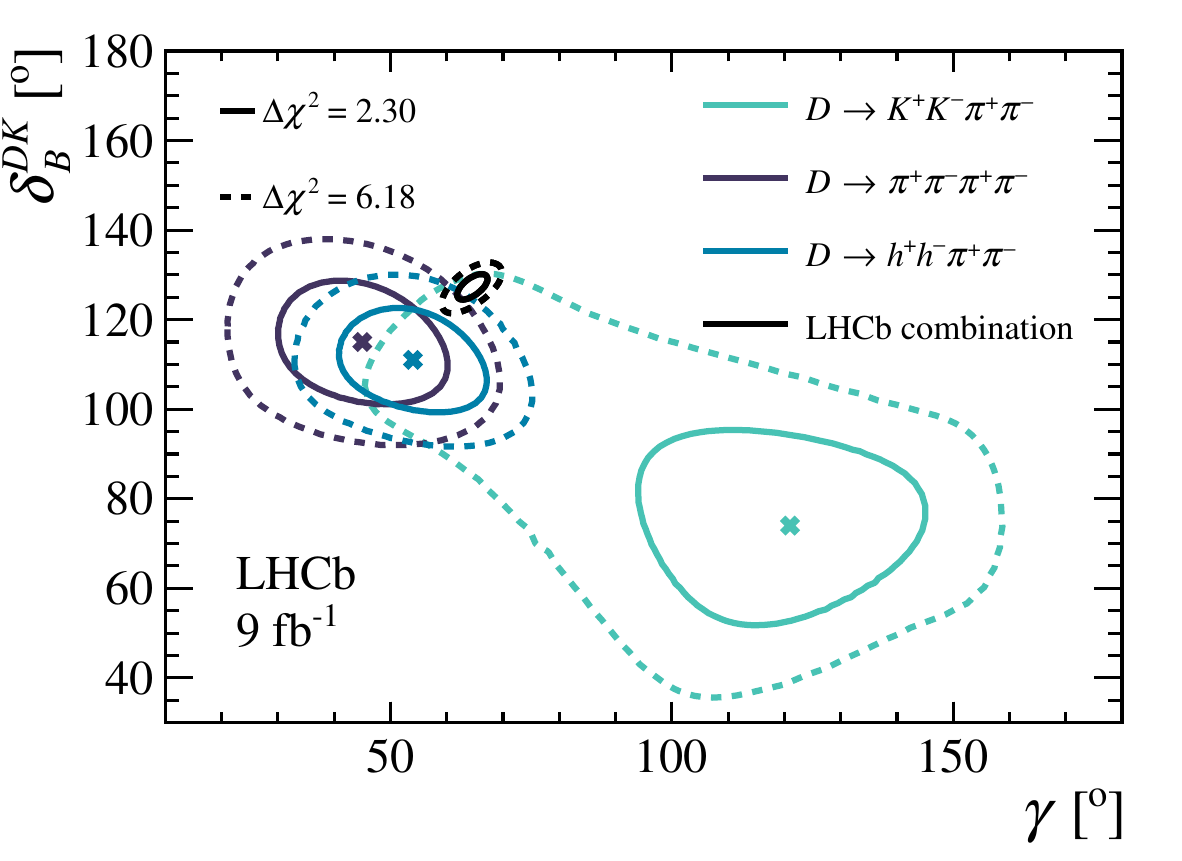}
    \end{subfigure}%
    \begin{subfigure}{0.5\textwidth}
        \includegraphics[width=1\textwidth]{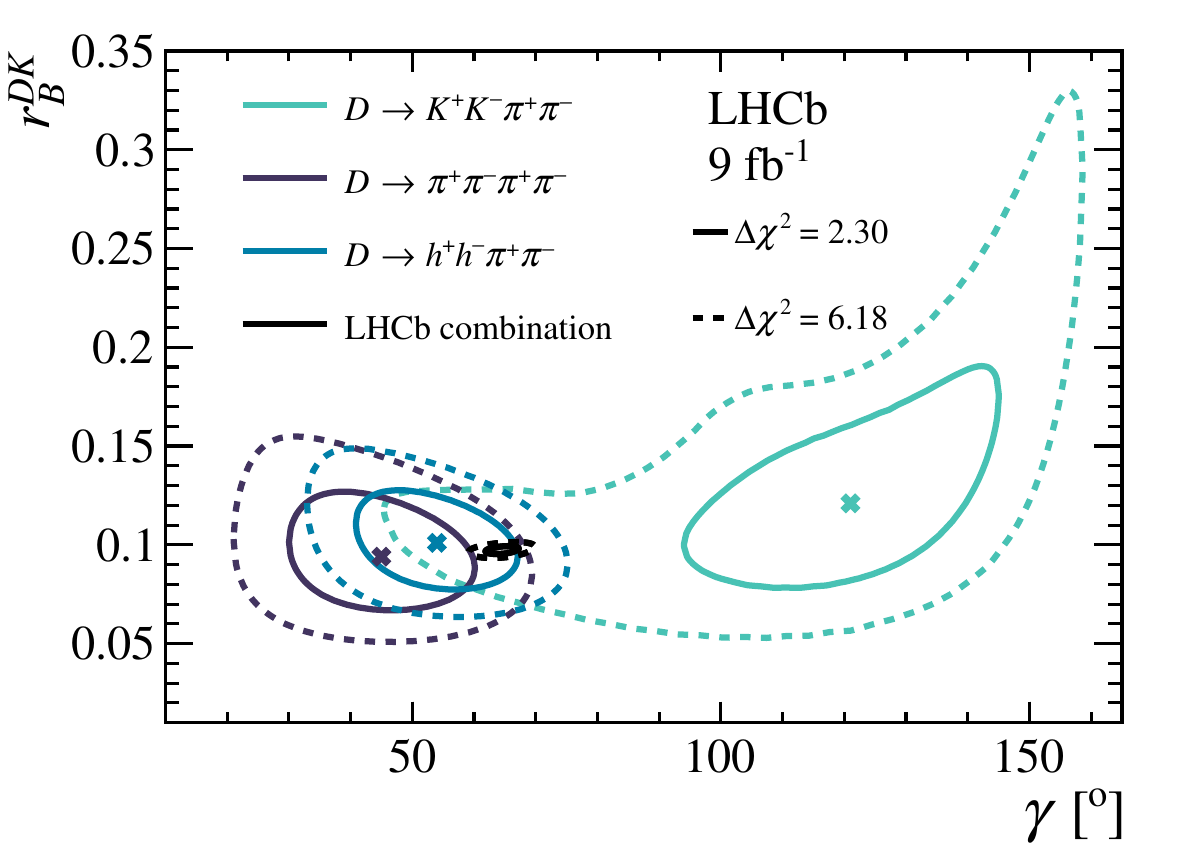}
    \end{subfigure}
    \caption{Interpretation of the phase-space binned measurement in terms to the underlying physics parameters. The variables $(\gamma, \delta_\B^{\D\kaon})$ are shown on the left and $(\gamma, r_\B^{\D\kaon})$ on the right. The 1$\sigma$ and 2$\sigma$ contours are shown with solid and dashed lines, which correspond to $\Delta\chi^2 = 2.30$ and $\Delta\chi^2 = 6.18$, respectively. The teal and purple contours show results from fitting the $\D\to\Kp\Km\pip\pim$ and $\pip\pim\pip\pim$ decays, respectively, and the blue contour is the result of a simultaneous fit of both modes. Also shown, in black, is the result of a combination of other decay modes performed by the \lhcb experiment~\cite{LHCb-CONF-2024-004}.}
    \label{figure:Prob_scan}
\end{figure}

In Fig.~\ref{figure:Prob_scan}, contours corresponding to $\Delta\chi^2 = 2.30$ and $\Delta\chi^2 = 6.18$ are shown. For variables that follow a Gaussian distribution, the contours are elliptical in shape. While this is approximately true for the contours of the $\D\to\pip\pim\pip\pim$ decay in purple, the teal contours describing the fit of $\D\to\Kp\Km\pip\pim$ show a highly non-Gaussian behaviour in the regions outside the 1$\sigma$ contour.

Such non-Gaussian behaviour is not seen when performing the interpretation with the $c_i$ and $s_i$ parameters fixed, nor was this reported in Ref.~\cite{LHCb-PAPER-2022-037}, where model-dependent inputs for these parameters were used. Thus, it may be concluded that it is the large uncertainties on these external charm strong-phase inputs, in particular those of $s_i$, that are driving this behaviour. Hence, it is not appropriate to assign a symmetric uncertainty on the physics parameters.

The black contours in Fig.~\ref{figure:Prob_scan} show the result from a combination of other decay modes performed by the \lhcb experiment~\cite{LHCb-CONF-2024-004}. When comparing the results from the $\D\to\Kp\Km\pip\pim$ mode in Fig.~\ref{figure:Prob_scan} with that reported in Ref.~\cite{LHCb-PAPER-2022-037}, the tension between the fit results of this decay mode and that of the \lhcb combination is significantly lower, once the model-independent uncertainties from $c_i$ and $s_i$ are accounted for. In addition, this illustrates the importance of determining the strong-phase differences of the charm decay with a model-independent method. Although a model may predict strong-phase parameters that are close to their true values, there currently exists no known strategy for evaluating their uncertainties accurately.

In the $\D\to\pip\pim\pip\pim$ channel, it is possible to run the fit with the binning scheme devised with CLEO-c data, which serves as a useful cross-check. Specifically, the optimal alternative scheme with $2\times3$ bins from Ref.~\cite{cite:4pi_cisi_CLEOc} shows the best sensitivity to $\gamma$. The bins and their corresponding values of $c_i$ and $s_i$ are all very distinct from the baseline binning by BESIII, but an excellent agreement in the fit results obtained from the two binning schemes is found. In particular, in a fit configuration where the strong-phase parameters $c_i$ and $s_i$ are fixed and without bias corrections, the fitted values of $\gamma$ from the BESIII and CLEO-c binning schemes are found to be $\gamma = (47 \pm 10)^\circ$ and $\gamma = (51 \pm 20)^\circ$, respectively. These uncertainties are statistical only, and the larger uncertainties from this alternate CLEO-c binning are expected from the central values of the $c_i$ and $s_i$ parameters.

\begin{figure}[tb]
    \centering
    \begin{subfigure}{0.5\textwidth}
        \includegraphics[width=1\textwidth]{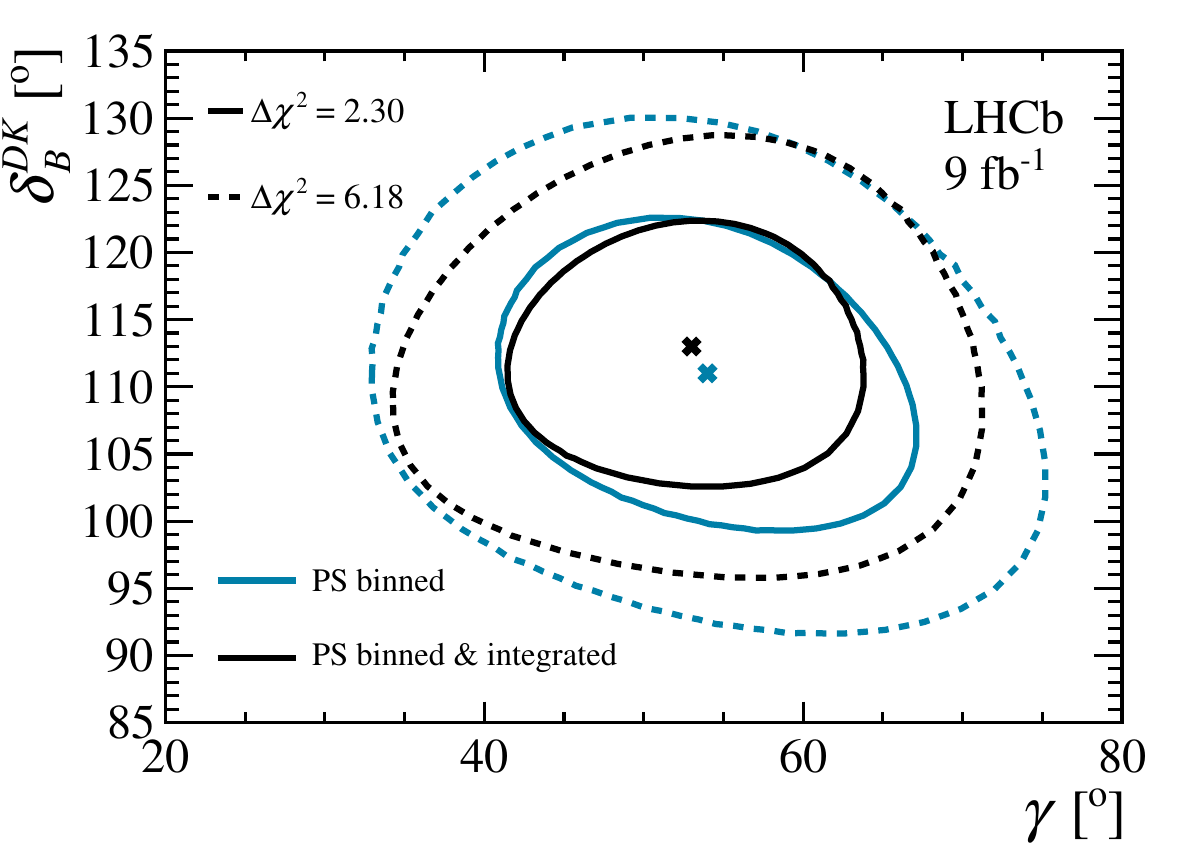}
    \end{subfigure}%
    \begin{subfigure}{0.5\textwidth}
        \includegraphics[width=1\textwidth]{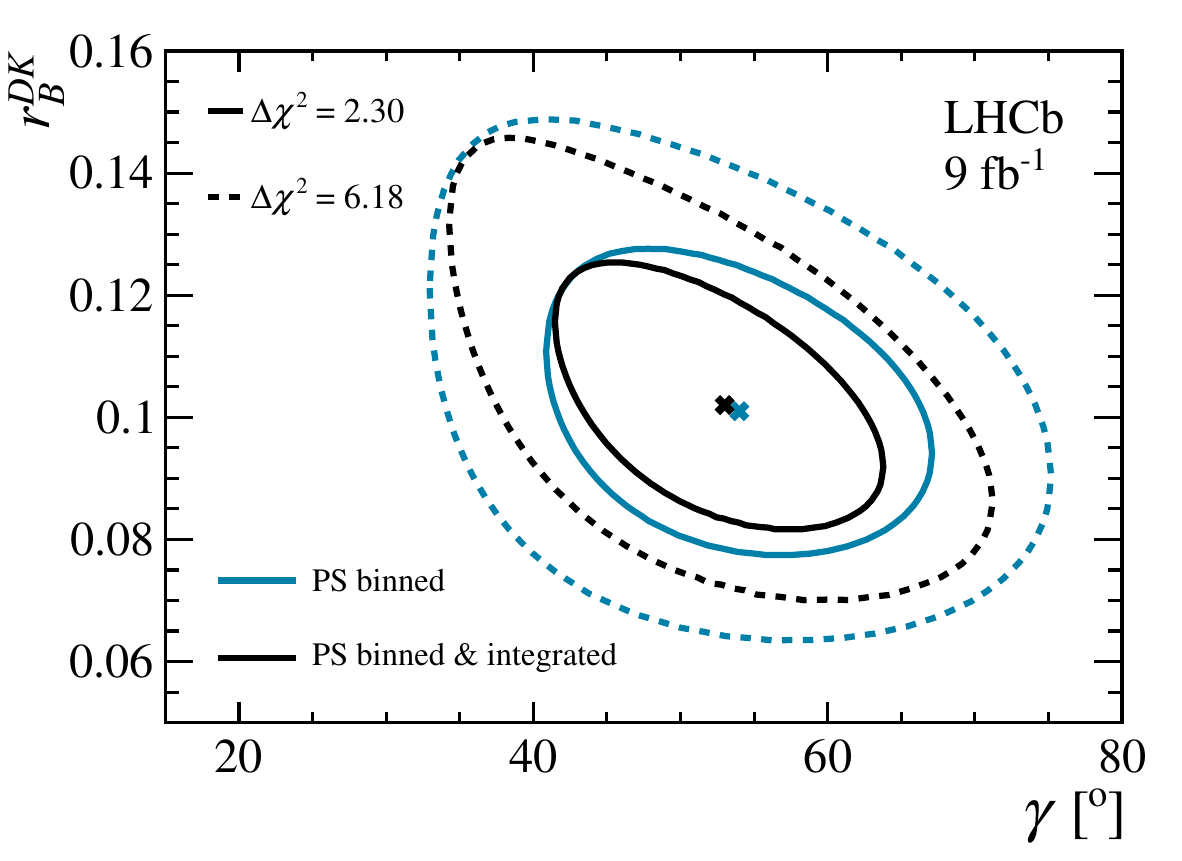}
    \end{subfigure}
    \caption{Contour comparison of the underlying physics parameters between the binned measurement, and the combined binned and integrated measurements. The 1$\sigma$ and 2$\sigma$ contours are shown with solid and dashed lines, which correspond to $\Delta\chi^2 = 2.30$ and $\Delta\chi^2 = 6.18$, respectively. The blue contour shows the results of the combined binned measurement of $\D\to\Kp\Km\pip\pim$ and $\pip\pim\pip\pim$, and the black contour also includes constraints from the phase-space integrated results from Ref.~\cite{LHCb-PAPER-2022-037}.}
    \label{figure:Prob_scan_GLW}
\end{figure}

It is also interesting to include information from the phase-space integrated \CP-violation effects, which were measured in Ref.~\cite{LHCb-PAPER-2022-037}. Furthermore, Ref.~\cite{LHCb-PAPER-2022-037} also established that the correlation between the phase-space binned and integrated \CP-violating observables can be neglected. Thus, a log-likelihood function is constructed using the central values and the covariance matrix from Ref.~\cite{LHCb-PAPER-2022-037}, assuming those \CP-violating observables follow a multidimensional Gaussian distribution. When adding the log-likelihood functions of the binned and integrated measurements, the minimisation results in the black contours shown in Fig.~\ref{figure:Prob_scan_GLW}. When comparing these black contours with those from the binned measurement, depicted by the blue lines, a small but non-negligible improvement is found.

A Feldman--Cousins approach using pseudoexperiments, referred to as the \textit{Plugin} method~\cite{Feldman-Cousins,cite:Plugin}, is used to assign statistically robust confidence intervals that correspond to a $68.3\%$ confidence level. This is to be compared with using Wilks' theorem, also referred to as the \textit{Prob} method, where the uncertainties tend to be underestimated, especially in cases where the statistical sensitivity is low or when fit parameters are near the edge of their allowed region.

\begin{figure}[tb]
    \centering
    \begin{subfigure}{0.5\textwidth}
        \includegraphics[width=1\textwidth]{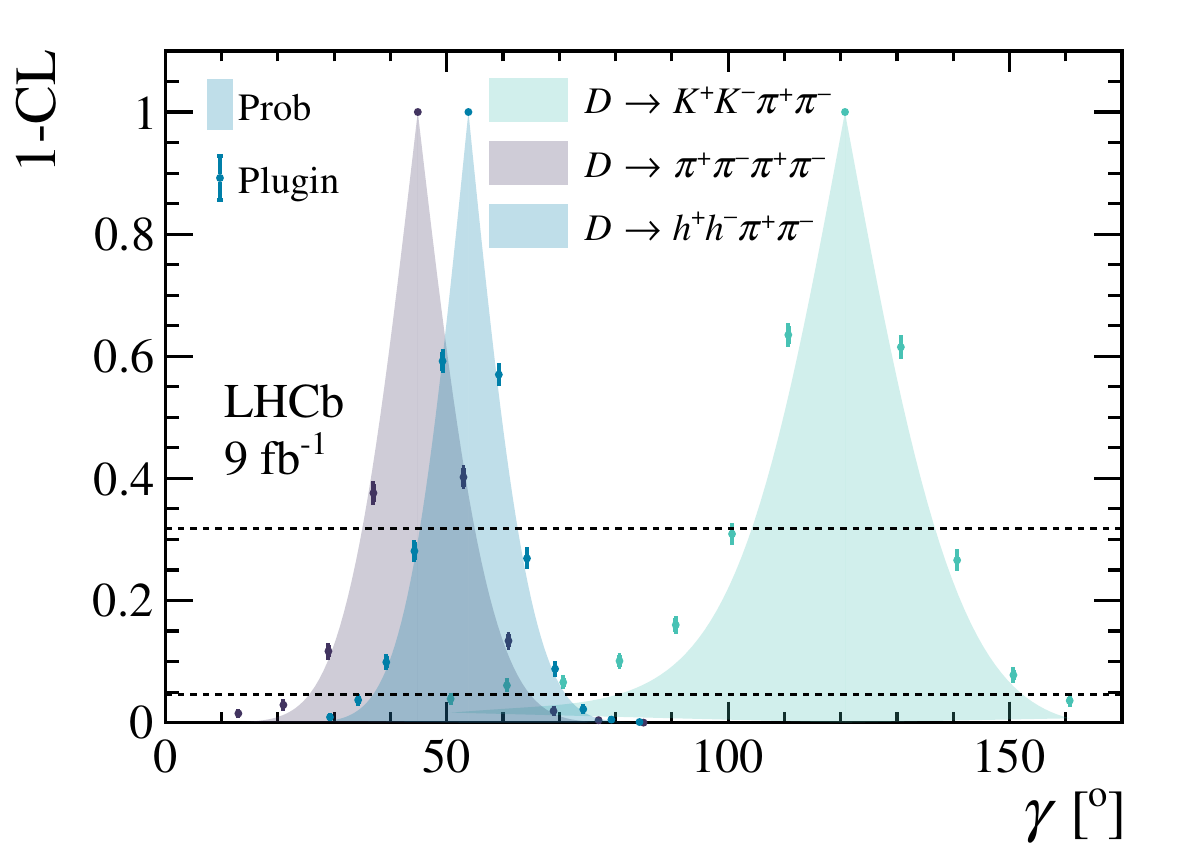}
    \end{subfigure}%
    \begin{subfigure}{0.5\textwidth}
        \includegraphics[width=1\textwidth]{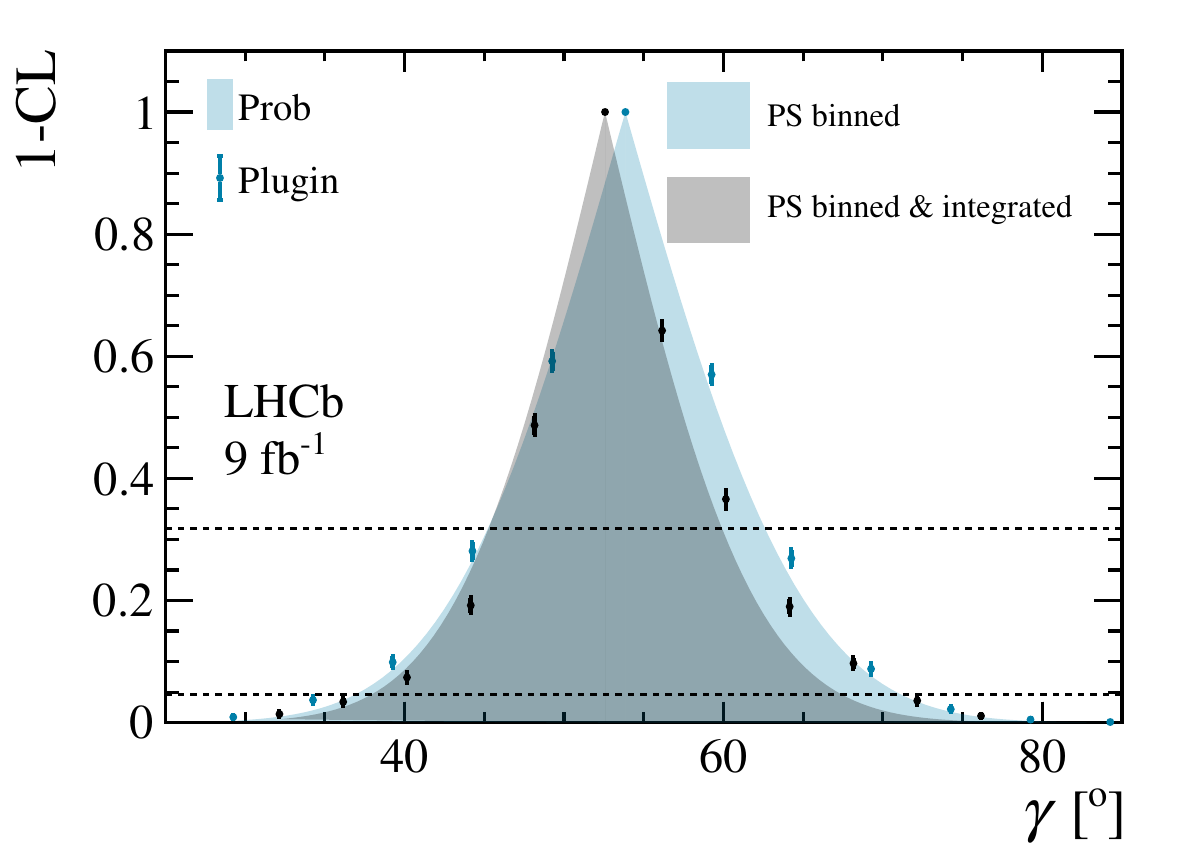}
    \end{subfigure}
    \caption{One-dimensional scans of the confidence level on the CKM angle $\gamma$. The solid lines are obtained using the Prob method. The data points are the results of the Plugin method and these confidence intervals are in general wider than those obtained from the Prob method. On the left is a comparison between fits to the (teal) $\D\to\Kp\Km\pip\pim$ and (purple) $\pip\pim\pip\pim$ channels, and (blue) the combined fit. On the right is a comparison between the binned measurement, and the combined binned and integrated measurements.}
    \label{figure:Plugin_scan}
\end{figure}

On the left of Fig.~\ref{figure:Plugin_scan}, the results from scans over the CKM angle $\gamma$ are shown as data points, using the same colour scheme as that of Fig.~\ref{figure:Prob_scan}. The solid shapes show the \textit{Prob} method for comparison. The confidence intervals can be read directly off the vertical axis, which shows $1 - \text{CL}$, where $\text{CL}$ denotes the confidence interval. The confidence intervals representing 1 and 2$\sigma$ are shown with the dashed lines. In general, the \textit{Prob} method yields a smaller uncertainty than the \textit{Plugin} method, particularly for the $\D\to\Kp\Km\pip\pim$ channel. Thus, the contours in Fig.~\ref{figure:Prob_scan} are in fact underestimated, which further reduces the tension between the two $\D$-decay modes. Similar scans are also performed over the other physics parameters, and the final results with both the $\D\to\Kp\Km\pip\pim$ and $\D\to\pip\pim\pip\pim$ modes combined, using the phase-space binned strategy, are
\begin{align*}
    \gamma =& (53.9_{-8.9}^{+9.5})^\circ, \\   
    \delta_\B^{\D\kaon} =& (111.0_{-7.9}^{+8.0})^\circ, \\
    r_\B^{\D\kaon} =& 0.101_{-0.018}^{+0.021}, \\
    \delta_\B^{\D\pion} =& (241_{-98}^{+103})^\circ, \\
    r_\B^{\D\pion} =& 0.0022_{-0.0022}^{+0.0041}.
\end{align*}

This result is the first binned model-independent measurement of $\gamma$ in the $\Bpm\to\D h^\pm$ decay, with the self-conjugate $\Kp\Km\pip\pim$ and $\pip\pim\pip\pim$ final states of the neutral charm meson. When combining with the phase-space integrated measurements of the same decays, the numerical values are
\begin{align*}
    \gamma =& (52.6_{-6.4}^{+8.5})^\circ, \\   
    \delta_\B^{\D\kaon} =& (112.6_{-7.8}^{+6.1})^\circ, \\
    r_\B^{\D\kaon} =& 0.102_{-0.017}^{+0.014}, \\
    \delta_\B^{\D\pion} =& (262_{-52}^{+40})^\circ, \\
    r_\B^{\D\pion} =& 0.0043_{-0.0043}^{+0.0033},
\end{align*}
making this one of the most precise measurements of $\gamma$ to date. The central value is also consistent with the combination of other \lhcb results~\cite{LHCb-CONF-2024-004}.

Several of the physics parameters have confidence intervals that are highly non-Gaussian distributed. In particular, the $r_\B^{\D\pion}$ parameter has a confidence interval that covers the whole region down to its physical boundary at zero. Such behaviour can be challenging when combining these results with other measurements of $\gamma$. Thus, for global combinations, an alternative parameterisation is provided in Appendix~\ref{appendix:Fit_results_with_a_Cartesian_parameterisation}.

\section{Summary and future prospects}
\label{section:Summary_and_future_prospects}
A first binned model-independent measurement of the CKM angle $\gamma$ has been performed using the $\Bpm\to\D h^\pm$ channel, with the subsequent four-body decays $\D\to\Kp\Km\pip\pim$ and $\D\to\pip\pim\pip\pim$. The measured value, using the phase-space binned approach, is $\gamma = (53.9_{-8.9}^{+9.5})^\circ$. A critical input to the analysis are the charm strong-phase parameters from \besiii~\cite{cite:KKpipi_cisi_BESIII,cite:4pi_cisi_BESIII}, which have recently become available. When combined with the phase-space integrated measurements from Ref.~\cite{LHCb-PAPER-2022-037}, the result is $\gamma = (52.6_{-6.4}^{+8.5})^\circ$. This is one of the most precise determinations of this SM parameter and is anticipated to make a significant impact on the global combination.

The measurement is currently statistically limited, and will benefit from the larger datasets that the \lhcb experiment is expected to collect with the \lhcb Upgrade I detector. Furthermore, the dominating systematic uncertainty is the contribution from strong-phase parameters, which can induce non-Gaussian behaviour in the distribution of $\gamma$ and other physics parameters. For the $\D\to\pip\pim\pip\pim$ channel, the current measurement from Ref.~\cite{cite:4pi_cisi_BESIII} can benefit from improved measurements with the full $20\invfb$ data set from \besiii. This is expected to reduce the statistical uncertainties on these parameters by approximately a factor $2.5$, ensuring that this remains a subdominant uncertainty on $\gamma$ and other physics parameters.

In the $\D\to\Kp\Km\pip\pim$ decay, where the strong-phase inputs use the full $20\invfb$ data set collected by the \besiii experiment, further improvements can be achieved through charm-mixing measurements by \lhcb. This will be crucial as the complementarity between the CKM angle $\gamma$ and charm mixing parameters can be exploited to improve the precision in both sectors.

\section*{Acknowledgements}
%
%
\noindent We express our gratitude to our colleagues in the CERN
accelerator departments for the excellent performance of the LHC. We
thank the technical and administrative staff at the LHCb
institutes.
We acknowledge support from CERN and from the national agencies:
ARC (Australia);
CAPES, CNPq, FAPERJ and FINEP (Brazil); 
MOST and NSFC (China); 
CNRS/IN2P3 (France); 
BMFTR, DFG and MPG (Germany); 
INFN (Italy); 
NWO (Netherlands); 
MNiSW and NCN (Poland); 
MCID/IFA (Romania); 
MICIU and AEI (Spain);
SNSF and SER (Switzerland); 
NASU (Ukraine); 
STFC (United Kingdom); 
DOE NP and NSF (USA).
We acknowledge the computing resources that are provided by ARDC (Australia), 
CBPF (Brazil),
CERN, 
IHEP and LZU (China),
IN2P3 (France), 
KIT and DESY (Germany), 
INFN (Italy), 
SURF (Netherlands),
Polish WLCG (Poland),
IFIN-HH (Romania), 
PIC (Spain), CSCS (Switzerland), 
and GridPP (United Kingdom).
We are indebted to the communities behind the multiple open-source
software packages on which we depend.
Individual groups or members have received support from
Key Research Program of Frontier Sciences of CAS, CAS PIFI, CAS CCEPP, 
Fundamental Research Funds for the Central Universities,  and Sci.\ \& Tech.\ Program of Guangzhou (China);
Minciencias (Colombia);
EPLANET, Marie Sk\l{}odowska-Curie Actions, ERC and NextGenerationEU (European Union);
A*MIDEX, ANR, IPhU and Labex P2IO, and R\'{e}gion Auvergne-Rh\^{o}ne-Alpes (France);
Alexander-von-Humboldt Foundation (Germany);
ICSC (Italy); 
Severo Ochoa and Mar\'ia de Maeztu Units of Excellence, GVA, XuntaGal, GENCAT, InTalent-Inditex and Prog.~Atracci\'on Talento CM (Spain);
SRC (Sweden);
the Leverhulme Trust, the Royal Society and UKRI (United Kingdom).



\section*{Appendices}

\appendix

\section{Fit results with a Cartesian parameterisation}
\label{appendix:Fit_results_with_a_Cartesian_parameterisation}
For global combinations, it may be desirable to use an alternative parameterisation that does not suffer from physical boundaries in the fit parameters. The reason is that large fit biases can be induced when the fitted parameters are close to a physical boundary, and the resulting uncertainties may not have the correct coverage. For this reason, it becomes challenging to combine this result with other measurements that are also sensitive to $\gamma$.

In previous \lhcb measurements using the same binned formalism~\cite{LHCb-PAPER-2024-023,LHCb-PAPER-2023-029,LHCb-PAPER-2023-012,LHCb-PAPER-2023-009,LHCb-PAPER-2020-019}, the strategy was to provide the results using the parameterisation with the six \CP-violating observables $x_\pm^{\D\kaon}$, $y_\pm^{\D\kaon}$, $x_\xi^{\D\pion}$ and $y_\xi^{\D\pion}$. However, as discussed in Sect.~\ref{section:Invariant_mass_fit}, the large uncertainties on $s_i$ results in uncertainties that do not follow a Gaussian distribution. Specifically, the $y_\pm^{\D\kaon}$ observables are found to have a significant asymmetry in their distributions, and it is therefore not suitable to use this parameterisation for global combinations.

In Ref.~\cite{LHCb-PAPER-2022-017}, a Cartesian parameterisation is proposed instead. This parameterisation uses $\gamma$ as a free parameter, as well as the Cartesian parameters $x^{\D h}$ and $y^{\D h}$, which are defined as
\begin{equation}
    x^{\D h} = r_\B^{\D h}\cos(\delta_\B^{\D h}), \quad y^{\D h} = r_\B^{\D h}\sin(\delta_\B^{\D h}),
\end{equation}
for $h = \kaon$ and $\pion$.

With the Cartesian parameterisation, the resulting fit parameters are shown in Table~\ref{table:Interpretation_fit_results_Cartesian}, and the corresponding correlation matrix is listed in Table~\ref{table:Interpretation_fit_correlations_Cartesian}. The values have been corrected for small fit biases, but the uncertainties are found to be well described using a multidimensional Gaussian distribution. The results in Table~\ref{table:Interpretation_fit_results_Cartesian} are performed without constraints from the phase-space integrated measurements in Ref.~\cite{LHCb-PAPER-2022-037}.

\begin{table}[htb]
    \centering
    \caption{The fitted physics parameters with a Cartesian parameterisation.}
    \label{table:Interpretation_fit_results_Cartesian}
    \begin{tabular}{cc} 
        \toprule
        Physics parameter                   & Fit result                          \\
        \midrule
        $\gamma$                            & $(53.7 \pm 7.8)^\circ$ \\
        $x^{\D\kaon}$                       & $(-3.7 \pm 1.4)\times 10^{-2}$ \\
        $y^{\D\kaon}$                       & $(\phantom{-}9.5 \pm 1.4)\times 10^{-2}$ \\
        $x^{\D\pion}$                       & $(-1.1 \pm 3.0)\times 10^{-3}$ \\
        $y^{\D\pion}$                       & $(-3.2 \pm 3.1)\times 10^{-3}$ \\
        \bottomrule
    \end{tabular}
\end{table}

\begin{table}[htb]
    \centering
    \caption{Correlation matrix of physics parameters with a Cartesian parameterisation.}
    \label{table:Interpretation_fit_correlations_Cartesian}
    \begin{tabular}{ccccc} 
        \toprule
        $\gamma$ & $x^{\D\kaon}$ & $y^{\D\kaon}$ & $x^{\D\pion}$ & $y^{\D\pion}$ \\
        \midrule
        $\phantom{-}1.000$  & $\phantom{-}0.142$  & $-0.018$ & $\phantom{-}0.006$  & $-0.108$ \\
        $\phantom{-}0.142$  & $\phantom{-}1.000$  & $-0.084$ & $-0.042$ & $\phantom{-}0.027$  \\
        $-0.018$ & $-0.084$ & $\phantom{-}1.000$  & $\phantom{-}0.045$  & $-0.072$ \\
        $\phantom{-}0.006$  & $-0.042$ & $\phantom{-}0.045$  & $\phantom{-}1.000$  & $-0.054$ \\
        $-0.108$ & $\phantom{-}0.027$  & $-0.072$ & $-0.054$ & $\phantom{-}1.000$  \\
        \bottomrule
    \end{tabular}
\end{table}

\newpage


\addcontentsline{toc}{section}{References}
\bibliographystyle{LHCb}
\bibliography{main,standard,LHCb-PAPER,LHCb-CONF,LHCb-DP,LHCb-TDR}

\ifx\mcitethebibliography\mciteundefinedmacro
\PackageError{LHCb.bst}{mciteplus.sty has not been loaded}
{This bibstyle requires the use of the mciteplus package.}\fi
\providecommand{\href}[2]{#2}
\begin{mcitethebibliography}{10}
\mciteSetBstSublistMode{n}
\mciteSetBstMaxWidthForm{subitem}{\alph{mcitesubitemcount})}
\mciteSetBstSublistLabelBeginEnd{\mcitemaxwidthsubitemform\space}
{\relax}{\relax}

\bibitem{cite:Sakharov}
A.~D. Sakharov, \ifthenelse{\boolean{articletitles}}{\emph{{Violation of CP invariance, C asymmetry, and baryon asymmetry of the universe}}, }{}\href{https://doi.org/10.1070/PU1991v034n05ABEH002497}{Pisma Zh.\ Eksp.\ Teor.\ Fiz.\  \textbf{5} (1967) 32}\relax
\mciteBstWouldAddEndPuncttrue
\mciteSetBstMidEndSepPunct{\mcitedefaultmidpunct}
{\mcitedefaultendpunct}{\mcitedefaultseppunct}\relax
\EndOfBibitem
\bibitem{Cabibbo:1963yz}
N.~Cabibbo, \ifthenelse{\boolean{articletitles}}{\emph{{Unitary symmetry and leptonic decays}}, }{}\href{https://doi.org/10.1103/PhysRevLett.10.531}{Phys.\ Rev.\ Lett.\  \textbf{10} (1963) 531}\relax
\mciteBstWouldAddEndPuncttrue
\mciteSetBstMidEndSepPunct{\mcitedefaultmidpunct}
{\mcitedefaultendpunct}{\mcitedefaultseppunct}\relax
\EndOfBibitem
\bibitem{Kobayashi:1973fv}
M.~Kobayashi and T.~Maskawa, \ifthenelse{\boolean{articletitles}}{\emph{{\CP-violation in the renormalizable theory of weak interaction}}, }{}\href{https://doi.org/10.1143/PTP.49.652}{Prog.\ Theor.\ Phys.\  \textbf{49} (1973) 652}\relax
\mciteBstWouldAddEndPuncttrue
\mciteSetBstMidEndSepPunct{\mcitedefaultmidpunct}
{\mcitedefaultendpunct}{\mcitedefaultseppunct}\relax
\EndOfBibitem
\bibitem{cite:Brod_Zupan}
J.~Brod and J.~Zupan, \ifthenelse{\boolean{articletitles}}{\emph{{The ultimate theoretical error on $\gamma$ from $B \to DK$ decays}}, }{}\href{https://doi.org/10.1007/JHEP01(2014)051}{JHEP \textbf{01} (2014) 051}, \href{http://arxiv.org/abs/1308.5663}{{\normalfont\ttfamily arXiv:1308.5663}}\relax
\mciteBstWouldAddEndPuncttrue
\mciteSetBstMidEndSepPunct{\mcitedefaultmidpunct}
{\mcitedefaultendpunct}{\mcitedefaultseppunct}\relax
\EndOfBibitem
\bibitem{Brod:2014bfa}
J.~Brod, A.~Lenz, G.~Tetlalmatzi-Xolocotzi, and M.~Wiebusch, \ifthenelse{\boolean{articletitles}}{\emph{{New physics effects in tree-level decays and the precision in the determination of the quark mixing angle {\ensuremath{\gamma}}}}, }{}\href{https://doi.org/10.1103/PhysRevD.92.033002}{Phys.\ Rev.\  \textbf{D92} (2015) 033002}, \href{http://arxiv.org/abs/1412.1446}{{\normalfont\ttfamily arXiv:1412.1446}}\relax
\mciteBstWouldAddEndPuncttrue
\mciteSetBstMidEndSepPunct{\mcitedefaultmidpunct}
{\mcitedefaultendpunct}{\mcitedefaultseppunct}\relax
\EndOfBibitem
\bibitem{Lenz:2019lvd}
A.~Lenz and G.~Tetlalmatzi-Xolocotzi, \ifthenelse{\boolean{articletitles}}{\emph{{Model-independent bounds on new physics effects in non-leptonic tree-level decays of B-mesons}}, }{}\href{https://doi.org/10.1007/JHEP07(2020)177}{JHEP \textbf{07} (2020) 177}, \href{http://arxiv.org/abs/1912.07621}{{\normalfont\ttfamily arXiv:1912.07621}}\relax
\mciteBstWouldAddEndPuncttrue
\mciteSetBstMidEndSepPunct{\mcitedefaultmidpunct}
{\mcitedefaultendpunct}{\mcitedefaultseppunct}\relax
\EndOfBibitem
\bibitem{CKMfitter2015}
CKMfitter group, J.~Charles {\em et~al.}, \ifthenelse{\boolean{articletitles}}{\emph{{Current status of the standard model CKM fit and constraints on \hbox{$\Delta F=2$} new physics}}, }{}\href{https://doi.org/10.1103/PhysRevD.91.073007}{Phys.\ Rev.\  \textbf{D91} (2015) 073007}, \href{http://arxiv.org/abs/1501.05013}{{\normalfont\ttfamily arXiv:1501.05013}}, {updated results and plots available at \href{http://ckmfitter.in2p3.fr/}{{\texttt{http://ckmfitter.in2p3.fr/}}}}\relax
\mciteBstWouldAddEndPuncttrue
\mciteSetBstMidEndSepPunct{\mcitedefaultmidpunct}
{\mcitedefaultendpunct}{\mcitedefaultseppunct}\relax
\EndOfBibitem
\bibitem{cite:UTfit}
UTfit collaboration, M.~Bona {\em et~al.}, \ifthenelse{\boolean{articletitles}}{\emph{{New UTfit analysis of the unitarity triangle in the Cabibbo--Kobayashi--Maskawa scheme}}, }{}\href{https://doi.org/10.1007/s12210-023-01137-5}{Rend.\ Lincei Sci.\ Fis.\ Nat.\  \textbf{34} (2023) 37}, \href{http://arxiv.org/abs/2212.03894}{{\normalfont\ttfamily arXiv:2212.03894}}\relax
\mciteBstWouldAddEndPuncttrue
\mciteSetBstMidEndSepPunct{\mcitedefaultmidpunct}
{\mcitedefaultendpunct}{\mcitedefaultseppunct}\relax
\EndOfBibitem
\bibitem{LHCb-CONF-2024-004}
{LHCb collaboration}, \ifthenelse{\boolean{articletitles}}{\emph{{Simultaneous determination of the CKM angle $\gamma$ and parameters related to mixing and \CP violation in the charm sector}}, }{} \href{http://cdsweb.cern.ch/search?p=LHCb-CONF-2024-004&f=reportnumber&action_search=Search&c=LHCb+Conference+Contributions} {LHCb-CONF-2024-004}, {2024}\relax
\mciteBstWouldAddEndPuncttrue
\mciteSetBstMidEndSepPunct{\mcitedefaultmidpunct}
{\mcitedefaultendpunct}{\mcitedefaultseppunct}\relax
\EndOfBibitem
\bibitem{LHCb-PAPER-2020-019}
LHCb collaboration, R.~Aaij {\em et~al.}, \ifthenelse{\boolean{articletitles}}{\emph{{Measurement of the CKM angle $\gamma$ in \mbox{$B^{\pm} \to D K^{\pm}$ and $B^{\pm} \to D \pi^{\pm}$} decays with $D \to \KS h^+h^-$}}, }{}\href{https://doi.org/10.1007/JHEP02(2021)169}{JHEP \textbf{02} (2021) 0169}, \href{http://arxiv.org/abs/2010.08483}{{\normalfont\ttfamily arXiv:2010.08483}}\relax
\mciteBstWouldAddEndPuncttrue
\mciteSetBstMidEndSepPunct{\mcitedefaultmidpunct}
{\mcitedefaultendpunct}{\mcitedefaultseppunct}\relax
\EndOfBibitem
\bibitem{cite:KShh_cisi_CLEO}
CLEO collaboration, J.~Libby {\em et~al.}, \ifthenelse{\boolean{articletitles}}{\emph{{Model-independent determination of the strong-phase difference between $\Dz$ and $\Dzb \to K^0_{S,L} h^+ h^-$ ($h=\pi,K$) and its impact on the measurement of the CKM angle $\gamma/\phi_3$}}, }{}\href{https://doi.org/10.1103/PhysRevD.82.112006}{Phys.\ Rev.\  \textbf{D82} (2010) 112006}, \href{http://arxiv.org/abs/1010.2817}{{\normalfont\ttfamily arXiv:1010.2817}}\relax
\mciteBstWouldAddEndPuncttrue
\mciteSetBstMidEndSepPunct{\mcitedefaultmidpunct}
{\mcitedefaultendpunct}{\mcitedefaultseppunct}\relax
\EndOfBibitem
\bibitem{cite:KSpipi_cisi_PRL}
BESIII collaboration, M.~Ablikim {\em et~al.}, \ifthenelse{\boolean{articletitles}}{\emph{{Determination of strong-phase parameters in $\D\rightarrow K^0_{S,L}\pi^+\pi^-$}}, }{}\href{https://doi.org/10.1103/PhysRevLett.124.241802}{Phys.\ Rev.\ Lett.\  \textbf{124} (2020) 241802}, \href{http://arxiv.org/abs/2002.12791}{{\normalfont\ttfamily arXiv:2002.12791}}\relax
\mciteBstWouldAddEndPuncttrue
\mciteSetBstMidEndSepPunct{\mcitedefaultmidpunct}
{\mcitedefaultendpunct}{\mcitedefaultseppunct}\relax
\EndOfBibitem
\bibitem{cite:KSpipi_cisi_PRD}
BESIII collaboration, M.~Ablikim {\em et~al.}, \ifthenelse{\boolean{articletitles}}{\emph{{Model-independent determination of the relative strong-phase difference between $\Dz$ and $\Dzb\rightarrow K^0_{S,L}\pi^+\pi^-$ and its impact on the measurement of the CKM angle $\gamma/\phi_3$}}, }{}\href{https://doi.org/10.1103/PhysRevD.101.112002}{Phys.\ Rev.\  \textbf{D101} (2020) 112002}, \href{http://arxiv.org/abs/2003.00091}{{\normalfont\ttfamily arXiv:2003.00091}}\relax
\mciteBstWouldAddEndPuncttrue
\mciteSetBstMidEndSepPunct{\mcitedefaultmidpunct}
{\mcitedefaultendpunct}{\mcitedefaultseppunct}\relax
\EndOfBibitem
\bibitem{cite:KSKK_cisi}
BESIII collaboration, M.~Ablikim {\em et~al.}, \ifthenelse{\boolean{articletitles}}{\emph{{Improved model-independent determination of the strong-phase difference between $\Dz$ and $\Dzb\to K^{0}_{\mathrm{S,L}}K^{+}K^{-}$ decays}}, }{}\href{https://doi.org/10.1103/PhysRevD.102.052008}{Phys.\ Rev.\  \textbf{D102} (2020) 052008}, \href{http://arxiv.org/abs/2007.07959}{{\normalfont\ttfamily arXiv:2007.07959}}\relax
\mciteBstWouldAddEndPuncttrue
\mciteSetBstMidEndSepPunct{\mcitedefaultmidpunct}
{\mcitedefaultendpunct}{\mcitedefaultseppunct}\relax
\EndOfBibitem
\bibitem{Rademacker:2006zx}
J.~Rademacker and G.~Wilkinson, \ifthenelse{\boolean{articletitles}}{\emph{{Determining the unitarity triangle $\gamma$ with a four-body amplitude analysis of $B^\pm\rightarrow(K^+ K^- \pi^+ \pi^-)_DK^\pm$ decays}}, }{}\href{https://doi.org/10.1016/j.physletb.2007.01.071}{Phys.\ Lett.\  \textbf{B647} (2007) 400}, \href{http://arxiv.org/abs/hep-ph/0611272}{{\normalfont\ttfamily arXiv:hep-ph/0611272}}\relax
\mciteBstWouldAddEndPuncttrue
\mciteSetBstMidEndSepPunct{\mcitedefaultmidpunct}
{\mcitedefaultendpunct}{\mcitedefaultseppunct}\relax
\EndOfBibitem
\bibitem{LHCb-PAPER-2022-037}
LHCb collaboration, R.~Aaij {\em et~al.}, \ifthenelse{\boolean{articletitles}}{\emph{{A study of CP violation in the decays \mbox{$\Bpm \rightarrow [\Kp\Km\pip\pim]_D \hadron^{\pm}$} \mbox{($\hadron=\kaon,\pion$)} and \mbox{$\Bpm \rightarrow [\pip\pim\pip\pim]_D \hadron^{\pm}$}}}, }{}\href{https://doi.org/10.1140/epjc/s10052-023-11560-5}{Eur.\ Phys.\ J.\  \textbf{C84} (2023) 547}, \href{http://arxiv.org/abs/2301.10328}{{\normalfont\ttfamily arXiv:2301.10328}}\relax
\mciteBstWouldAddEndPuncttrue
\mciteSetBstMidEndSepPunct{\mcitedefaultmidpunct}
{\mcitedefaultendpunct}{\mcitedefaultseppunct}\relax
\EndOfBibitem
\bibitem{LHCb-PAPER-2018-041}
LHCb collaboration, R.~Aaij {\em et~al.}, \ifthenelse{\boolean{articletitles}}{\emph{{Search for \CP violation through an amplitude analysis of \mbox{\decay{\Dz}{\Kp\Km\pip\pim}} decays}}, }{}\href{https://doi.org/10.1007/JHEP02(2019)126}{JHEP \textbf{02} (2019) 126}, \href{http://arxiv.org/abs/1811.08304}{{\normalfont\ttfamily arXiv:1811.08304}}\relax
\mciteBstWouldAddEndPuncttrue
\mciteSetBstMidEndSepPunct{\mcitedefaultmidpunct}
{\mcitedefaultendpunct}{\mcitedefaultseppunct}\relax
\EndOfBibitem
\bibitem{cite:KKpipi_cisi_BESIII}
BESIII collaboration, M.~Ablikim {\em et~al.}, \ifthenelse{\boolean{articletitles}}{\emph{{Measurement of the strong-phase difference between $\Dz$ and $\Dzb\to K^+K^-\pi^+\pi^-$ in bins of phase space}}, }{}\href{https://doi.org/10.1103/PhysRevD.112.012015}{Phys.\ Rev.\  \textbf{D112} (2025) 012015}, \href{http://arxiv.org/abs/2502.12873}{{\normalfont\ttfamily arXiv:2502.12873}}\relax
\mciteBstWouldAddEndPuncttrue
\mciteSetBstMidEndSepPunct{\mcitedefaultmidpunct}
{\mcitedefaultendpunct}{\mcitedefaultseppunct}\relax
\EndOfBibitem
\bibitem{cite:4pi_cisi_CLEOc}
S.~Harnew {\em et~al.}, \ifthenelse{\boolean{articletitles}}{\emph{{Model-independent determination of the strong phase difference between $\Dz$ and $\Dzb \to\pi^+\pi^-\pi^+\pi^-$ amplitudes}}, }{}\href{https://doi.org/10.1007/JHEP01(2018)144}{JHEP \textbf{01} (2018) 144}, \href{http://arxiv.org/abs/1709.03467}{{\normalfont\ttfamily arXiv:1709.03467}}\relax
\mciteBstWouldAddEndPuncttrue
\mciteSetBstMidEndSepPunct{\mcitedefaultmidpunct}
{\mcitedefaultendpunct}{\mcitedefaultseppunct}\relax
\EndOfBibitem
\bibitem{cite:4pi_model_BESIII}
BESIII collaboration, M.~Ablikim {\em et~al.}, \ifthenelse{\boolean{articletitles}}{\emph{{Amplitude analysis of the decays $\Dz\to\pi^+\pi^-\pi^+\pi^-$ and $\Dz\to\pi^+\pi^-\pi^0\pi^0$}}, }{}\href{https://doi.org/10.1088/1674-1137/ad3d4d}{Chin.\ Phys.\  \textbf{C48} (2024) 083001}, \href{http://arxiv.org/abs/2312.02524}{{\normalfont\ttfamily arXiv:2312.02524}}\relax
\mciteBstWouldAddEndPuncttrue
\mciteSetBstMidEndSepPunct{\mcitedefaultmidpunct}
{\mcitedefaultendpunct}{\mcitedefaultseppunct}\relax
\EndOfBibitem
\bibitem{cite:4pi_cisi_BESIII}
BESIII collaboration, M.~Ablikim {\em et~al.}, \ifthenelse{\boolean{articletitles}}{\emph{{Model-independent determination of the strong-phase difference between $\Dz$ and $\Dzb\to\pi^+\pi^-\pi^+\pi^-$ decays}}, }{}\href{https://doi.org/10.1103/PhysRevD.110.112008}{Phys.\ Rev.\  \textbf{D110} (2024) 112008}, \href{http://arxiv.org/abs/2408.16279}{{\normalfont\ttfamily arXiv:2408.16279}}\relax
\mciteBstWouldAddEndPuncttrue
\mciteSetBstMidEndSepPunct{\mcitedefaultmidpunct}
{\mcitedefaultendpunct}{\mcitedefaultseppunct}\relax
\EndOfBibitem
\bibitem{cite:CharmCPVgamma}
A.~Bondar, A.~Dolgov, A.~Poluektov, and V.~Vorobiev, \ifthenelse{\boolean{articletitles}}{\emph{{Effect of direct CP violation in charm on $\gamma$ extraction from $B^\pm \to DK^{\pm}, D \to K^0_S \pi^+ \pi^-$ Dalitz plot analysis}}, }{}\href{https://doi.org/10.1140/epjc/s10052-013-2476-9}{Eur.\ Phys.\ J.\  \textbf{C73} (2013) 2476}, \href{http://arxiv.org/abs/1303.6305}{{\normalfont\ttfamily arXiv:1303.6305}}\relax
\mciteBstWouldAddEndPuncttrue
\mciteSetBstMidEndSepPunct{\mcitedefaultmidpunct}
{\mcitedefaultendpunct}{\mcitedefaultseppunct}\relax
\EndOfBibitem
\bibitem{PDG2024}
Particle Data Group, S.~Navas {\em et~al.}, \ifthenelse{\boolean{articletitles}}{\emph{{\href{http://pdg.lbl.gov/}{Review of particle physics}}}, }{}\href{https://doi.org/10.1103/PhysRevD.110.030001}{Phys.\ Rev.\  \textbf{D110} (2024) 030001}\relax
\mciteBstWouldAddEndPuncttrue
\mciteSetBstMidEndSepPunct{\mcitedefaultmidpunct}
{\mcitedefaultendpunct}{\mcitedefaultseppunct}\relax
\EndOfBibitem
\bibitem{cite:Simultaneous_B_gamma_fit}
J.~Garra~Tic\'o {\em et~al.}, \ifthenelse{\boolean{articletitles}}{\emph{{Study of the sensitivity to CKM angle $\gamma$ under simultaneous determination from multiple $B$ meson decay modes}}, }{}\href{https://doi.org/10.1103/PhysRevD.102.053003}{Phys.\ Rev.\  \textbf{D102} (2020) 053003}, \href{http://arxiv.org/abs/1909.00600}{{\normalfont\ttfamily arXiv:1909.00600}}\relax
\mciteBstWouldAddEndPuncttrue
\mciteSetBstMidEndSepPunct{\mcitedefaultmidpunct}
{\mcitedefaultendpunct}{\mcitedefaultseppunct}\relax
\EndOfBibitem
\bibitem{LHCb-PAPER-2022-017}
LHCb collaboration, R.~Aaij {\em et~al.}, \ifthenelse{\boolean{articletitles}}{\emph{{Measurement of the CKM angle $\gamma$ with \mbox{$\Bmp \rightarrow D[\Kpm\pimp\pimp\pipm] h^{\mp}$ decays using a binned phase-space approach}}}, }{}\href{https://doi.org/10.1007/JHEP07(2023)138}{JHEP \textbf{07} (2023) 138}, \href{http://arxiv.org/abs/2209.03692}{{\normalfont\ttfamily arXiv:2209.03692}}\relax
\mciteBstWouldAddEndPuncttrue
\mciteSetBstMidEndSepPunct{\mcitedefaultmidpunct}
{\mcitedefaultendpunct}{\mcitedefaultseppunct}\relax
\EndOfBibitem
\bibitem{Rama:2013voa}
M.~Rama, \ifthenelse{\boolean{articletitles}}{\emph{{Effect of $\D-\Db$ mixing in the extraction of $\gamma$ with $\Bm\to\Dz\Km$ and $\Bm\to\Dz\pim$ decays}}, }{}\href{https://doi.org/10.1103/PhysRevD.89.014021}{Phys.\ Rev.\  \textbf{D89} (2014) 014021}, \href{http://arxiv.org/abs/1307.4384}{{\normalfont\ttfamily arXiv:1307.4384}}\relax
\mciteBstWouldAddEndPuncttrue
\mciteSetBstMidEndSepPunct{\mcitedefaultmidpunct}
{\mcitedefaultendpunct}{\mcitedefaultseppunct}\relax
\EndOfBibitem
\bibitem{HFLAV21}
Heavy Flavor Averaging Group, Y.~Amhis {\em et~al.}, \ifthenelse{\boolean{articletitles}}{\emph{{Averages of $b$-hadron, $c$-hadron, and $\tau$-lepton properties as of 2021}}, }{}\href{https://doi.org/10.1103/PhysRevD.107.052008}{Phys.\ Rev.\  \textbf{D107} (2023) 052008}, \href{http://arxiv.org/abs/2206.07501}{{\normalfont\ttfamily arXiv:2206.07501}}, {updated results and plots available at \href{https://hflav.web.cern.ch}{{\texttt{https://hflav.web.cern.ch}}}}\relax
\mciteBstWouldAddEndPuncttrue
\mciteSetBstMidEndSepPunct{\mcitedefaultmidpunct}
{\mcitedefaultendpunct}{\mcitedefaultseppunct}\relax
\EndOfBibitem
\bibitem{cite:KSpipipi0_cisi}
P.~K. Resmi, J.~Libby, S.~Malde, and G.~Wilkinson, \ifthenelse{\boolean{articletitles}}{\emph{{Quantum-correlated measurements of $\D\to K^{0}_{\rm S}\pi^{+}\pi^{-}\pi^{0}$ decays and consequences for the determination of the CKM angle $\gamma$}}, }{}\href{https://doi.org/10.1007/JHEP01(2018)082}{JHEP \textbf{01} (2018) 082}, \href{http://arxiv.org/abs/1710.10086}{{\normalfont\ttfamily arXiv:1710.10086}}\relax
\mciteBstWouldAddEndPuncttrue
\mciteSetBstMidEndSepPunct{\mcitedefaultmidpunct}
{\mcitedefaultendpunct}{\mcitedefaultseppunct}\relax
\EndOfBibitem
\bibitem{cite:Kpipipi_deltaD}
T.~Evans, J.~Libby, S.~Malde, and G.~Wilkinson, \ifthenelse{\boolean{articletitles}}{\emph{{Improved sensitivity to the CKM phase $\gamma$ through binning phase space in $B^- \to \D K^-$, $\D \to K^+\pi^-\pi^-\pi^+$ decays}}, }{}\href{https://doi.org/10.1016/j.physletb.2019.135188}{Phys.\ Lett.\  \textbf{B802} (2020) 135188}, \href{http://arxiv.org/abs/1909.10196}{{\normalfont\ttfamily arXiv:1909.10196}}\relax
\mciteBstWouldAddEndPuncttrue
\mciteSetBstMidEndSepPunct{\mcitedefaultmidpunct}
{\mcitedefaultendpunct}{\mcitedefaultseppunct}\relax
\EndOfBibitem
\bibitem{cite:KKpipiBinningScheme}
M.~Tat {\em et~al.}, \ifthenelse{\boolean{articletitles}}{\emph{{D0-\textgreater KKpipi binning scheme}}, }{} 2022.
\newblock doi:~\href{https://doi.org/10.5281/zenodo.6940031}{10.5281/zenodo.6940031}\relax
\mciteBstWouldAddEndPuncttrue
\mciteSetBstMidEndSepPunct{\mcitedefaultmidpunct}
{\mcitedefaultendpunct}{\mcitedefaultseppunct}\relax
\EndOfBibitem
\bibitem{cite:4pi_model_CLEOc}
P.~d'Argent {\em et~al.}, \ifthenelse{\boolean{articletitles}}{\emph{{Amplitude analyses of $\Dz \to {\pi^+\pi^-\pi^+\pi^-}$ and $\Dz \to {K^+K^-\pi^+\pi^-}$ decays}}, }{}\href{https://doi.org/10.1007/JHEP05(2017)143}{JHEP \textbf{05} (2017) 143}, \href{http://arxiv.org/abs/1703.08505}{{\normalfont\ttfamily arXiv:1703.08505}}\relax
\mciteBstWouldAddEndPuncttrue
\mciteSetBstMidEndSepPunct{\mcitedefaultmidpunct}
{\mcitedefaultendpunct}{\mcitedefaultseppunct}\relax
\EndOfBibitem
\bibitem{cite:4piBinningScheme}
X.~Shi {\em et~al.}, \ifthenelse{\boolean{articletitles}}{\emph{{The binning schemes of D0-\textgreater pi+pi+pi-pi-}}, }{} 2024.
\newblock doi:~\href{https://doi.org/10.5281/zenodo.14029568}{10.5281/zenodo.14029568}\relax
\mciteBstWouldAddEndPuncttrue
\mciteSetBstMidEndSepPunct{\mcitedefaultmidpunct}
{\mcitedefaultendpunct}{\mcitedefaultseppunct}\relax
\EndOfBibitem
\bibitem{LHCb-DP-2008-001}
LHCb collaboration, A.~A. Alves~Jr.\ {\em et~al.}, \ifthenelse{\boolean{articletitles}}{\emph{{The \lhcb detector at the LHC}}, }{}\href{https://doi.org/10.1088/1748-0221/3/08/S08005}{JINST \textbf{3} (2008) S08005}\relax
\mciteBstWouldAddEndPuncttrue
\mciteSetBstMidEndSepPunct{\mcitedefaultmidpunct}
{\mcitedefaultendpunct}{\mcitedefaultseppunct}\relax
\EndOfBibitem
\bibitem{LHCb-DP-2014-002}
LHCb collaboration, R.~Aaij {\em et~al.}, \ifthenelse{\boolean{articletitles}}{\emph{{LHCb detector performance}}, }{}\href{https://doi.org/10.1142/S0217751X15300227}{Int.\ J.\ Mod.\ Phys.\  \textbf{A30} (2015) 1530022}, \href{http://arxiv.org/abs/1412.6352}{{\normalfont\ttfamily arXiv:1412.6352}}\relax
\mciteBstWouldAddEndPuncttrue
\mciteSetBstMidEndSepPunct{\mcitedefaultmidpunct}
{\mcitedefaultendpunct}{\mcitedefaultseppunct}\relax
\EndOfBibitem
\bibitem{Sjostrand:2007gs}
T.~Sj\"{o}strand, S.~Mrenna, and P.~Skands, \ifthenelse{\boolean{articletitles}}{\emph{{A brief introduction to PYTHIA 8.1}}, }{}\href{https://doi.org/10.1016/j.cpc.2008.01.036}{Comput.\ Phys.\ Commun.\  \textbf{178} (2008) 852}, \href{http://arxiv.org/abs/0710.3820}{{\normalfont\ttfamily arXiv:0710.3820}}\relax
\mciteBstWouldAddEndPuncttrue
\mciteSetBstMidEndSepPunct{\mcitedefaultmidpunct}
{\mcitedefaultendpunct}{\mcitedefaultseppunct}\relax
\EndOfBibitem
\bibitem{Sjostrand:2006za}
T.~Sj\"{o}strand, S.~Mrenna, and P.~Skands, \ifthenelse{\boolean{articletitles}}{\emph{{PYTHIA 6.4 physics and manual}}, }{}\href{https://doi.org/10.1088/1126-6708/2006/05/026}{JHEP \textbf{05} (2006) 026}, \href{http://arxiv.org/abs/hep-ph/0603175}{{\normalfont\ttfamily arXiv:hep-ph/0603175}}\relax
\mciteBstWouldAddEndPuncttrue
\mciteSetBstMidEndSepPunct{\mcitedefaultmidpunct}
{\mcitedefaultendpunct}{\mcitedefaultseppunct}\relax
\EndOfBibitem
\bibitem{LHCb-PROC-2010-056}
I.~Belyaev {\em et~al.}, \ifthenelse{\boolean{articletitles}}{\emph{{Handling of the generation of primary events in Gauss, the LHCb simulation framework}}, }{}\href{https://doi.org/10.1088/1742-6596/331/3/032047}{J.\ Phys.\ Conf.\ Ser.\  \textbf{331} (2011) 032047}\relax
\mciteBstWouldAddEndPuncttrue
\mciteSetBstMidEndSepPunct{\mcitedefaultmidpunct}
{\mcitedefaultendpunct}{\mcitedefaultseppunct}\relax
\EndOfBibitem
\bibitem{Lange:2001uf}
D.~J. Lange, \ifthenelse{\boolean{articletitles}}{\emph{{The EvtGen particle decay simulation package}}, }{}\href{https://doi.org/10.1016/S0168-9002(01)00089-4}{Nucl.\ Instrum.\ Meth.\  \textbf{A462} (2001) 152}\relax
\mciteBstWouldAddEndPuncttrue
\mciteSetBstMidEndSepPunct{\mcitedefaultmidpunct}
{\mcitedefaultendpunct}{\mcitedefaultseppunct}\relax
\EndOfBibitem
\bibitem{davidson2015photos}
N.~Davidson, T.~Przedzinski, and Z.~Was, \ifthenelse{\boolean{articletitles}}{\emph{{PHOTOS interface in C++: Technical and physics documentation}}, }{}\href{https://doi.org/https://doi.org/10.1016/j.cpc.2015.09.013}{Comput.\ Phys.\ Commun.\  \textbf{199} (2016) 86}, \href{http://arxiv.org/abs/1011.0937}{{\normalfont\ttfamily arXiv:1011.0937}}\relax
\mciteBstWouldAddEndPuncttrue
\mciteSetBstMidEndSepPunct{\mcitedefaultmidpunct}
{\mcitedefaultendpunct}{\mcitedefaultseppunct}\relax
\EndOfBibitem
\bibitem{Allison:2006ve}
Geant4 collaboration, J.~Allison {\em et~al.}, \ifthenelse{\boolean{articletitles}}{\emph{{Geant4 developments and applications}}, }{}\href{https://doi.org/10.1109/TNS.2006.869826}{IEEE Trans.\ Nucl.\ Sci.\  \textbf{53} (2006) 270}\relax
\mciteBstWouldAddEndPuncttrue
\mciteSetBstMidEndSepPunct{\mcitedefaultmidpunct}
{\mcitedefaultendpunct}{\mcitedefaultseppunct}\relax
\EndOfBibitem
\bibitem{Agostinelli:2002hh}
Geant4 collaboration, S.~Agostinelli {\em et~al.}, \ifthenelse{\boolean{articletitles}}{\emph{{Geant4: A simulation toolkit}}, }{}\href{https://doi.org/10.1016/S0168-9002(03)01368-8}{Nucl.\ Instrum.\ Meth.\  \textbf{A506} (2003) 250}\relax
\mciteBstWouldAddEndPuncttrue
\mciteSetBstMidEndSepPunct{\mcitedefaultmidpunct}
{\mcitedefaultendpunct}{\mcitedefaultseppunct}\relax
\EndOfBibitem
\bibitem{LHCb-PROC-2011-006}
M.~Clemencic {\em et~al.}, \ifthenelse{\boolean{articletitles}}{\emph{{The \lhcb simulation application, Gauss: Design, evolution and experience}}, }{}\href{https://doi.org/10.1088/1742-6596/331/3/032023}{J.\ Phys.\ Conf.\ Ser.\  \textbf{331} (2011) 032023}\relax
\mciteBstWouldAddEndPuncttrue
\mciteSetBstMidEndSepPunct{\mcitedefaultmidpunct}
{\mcitedefaultendpunct}{\mcitedefaultseppunct}\relax
\EndOfBibitem
\bibitem{LHCb-DP-2018-004}
D.~M{\"u}ller, M.~Clemencic, G.~Corti, and M.~Gersabeck, \ifthenelse{\boolean{articletitles}}{\emph{{ReDecay: A novel approach to speed up the simulation at LHCb}}, }{}\href{https://doi.org/10.1140/epjc/s10052-018-6469-6}{Eur.\ Phys.\ J.\  \textbf{C78} (2018) 1009}, \href{http://arxiv.org/abs/1810.10362}{{\normalfont\ttfamily arXiv:1810.10362}}\relax
\mciteBstWouldAddEndPuncttrue
\mciteSetBstMidEndSepPunct{\mcitedefaultmidpunct}
{\mcitedefaultendpunct}{\mcitedefaultseppunct}\relax
\EndOfBibitem
\bibitem{Breiman}
L.~Breiman, J.~H. Friedman, R.~A. Olshen, and C.~J. Stone, {\em Classification and regression trees}, Wadsworth international group, Belmont, California, USA, 1984\relax
\mciteBstWouldAddEndPuncttrue
\mciteSetBstMidEndSepPunct{\mcitedefaultmidpunct}
{\mcitedefaultendpunct}{\mcitedefaultseppunct}\relax
\EndOfBibitem
\bibitem{AdaBoost}
Y.~Freund and R.~E. Schapire, \ifthenelse{\boolean{articletitles}}{\emph{A decision-theoretic generalization of on-line learning and an application to boosting}, }{}\href{https://doi.org/10.1006/jcss.1997.1504}{J.\ Comput.\ Syst.\ Sci.\  \textbf{55} (1997) 119}\relax
\mciteBstWouldAddEndPuncttrue
\mciteSetBstMidEndSepPunct{\mcitedefaultmidpunct}
{\mcitedefaultendpunct}{\mcitedefaultseppunct}\relax
\EndOfBibitem
\bibitem{Hocker:2007ht}
H.~Voss, A.~Hoecker, J.~Stelzer, and F.~Tegenfeldt, \ifthenelse{\boolean{articletitles}}{\emph{{TMVA - Toolkit for Multivariate Data Analysis with ROOT}}, }{}\href{https://doi.org/10.22323/1.050.0040}{PoS \textbf{ACAT} (2007) 040}\relax
\mciteBstWouldAddEndPuncttrue
\mciteSetBstMidEndSepPunct{\mcitedefaultmidpunct}
{\mcitedefaultendpunct}{\mcitedefaultseppunct}\relax
\EndOfBibitem
\bibitem{TMVA4}
A.~Hoecker {\em et~al.}, \ifthenelse{\boolean{articletitles}}{\emph{{TMVA 4 --- Toolkit for Multivariate Data Analysis with ROOT. Users Guide.}}, }{}\href{http://arxiv.org/abs/physics/0703039}{{\normalfont\ttfamily arXiv:physics/0703039}}\relax
\mciteBstWouldAddEndPuncttrue
\mciteSetBstMidEndSepPunct{\mcitedefaultmidpunct}
{\mcitedefaultendpunct}{\mcitedefaultseppunct}\relax
\EndOfBibitem
\bibitem{LHCb-PAPER-2018-017}
LHCb collaboration, R.~Aaij {\em et~al.}, \ifthenelse{\boolean{articletitles}}{\emph{{Measurement of the CKM angle $\gamma$ using \mbox{\decay{\Bpm}{D\Kpm}} with \mbox{\decay{D}{\KS \pip \pim,\ \KS \Kp\Km}} decays}}, }{}\href{https://doi.org/10.1007/JHEP08(2018)176}{JHEP \textbf{08} (2018) 176}, Erratum \href{https://doi.org/10.1007/JHEP10(2018)107}{ibid.\   \textbf{10} (2018) 107}, \href{http://arxiv.org/abs/1806.01202}{{\normalfont\ttfamily arXiv:1806.01202}}\relax
\mciteBstWouldAddEndPuncttrue
\mciteSetBstMidEndSepPunct{\mcitedefaultmidpunct}
{\mcitedefaultendpunct}{\mcitedefaultseppunct}\relax
\EndOfBibitem
\bibitem{LHCb-PAPER-2020-036}
LHCb collaboration, R.~Aaij {\em et~al.}, \ifthenelse{\boolean{articletitles}}{\emph{{Measurement of \CP observables in $B^\pm \to D^{(*)} K^{\pm}$ and $B^\pm \to D^{(*)} \pi^{\pm} $ decays using two-body $D$ final states}}, }{}\href{https://doi.org/10.1007/JHEP04(2021)081}{JHEP \textbf{04} (2021) 081}, \href{http://arxiv.org/abs/2012.09903}{{\normalfont\ttfamily arXiv:2012.09903}}\relax
\mciteBstWouldAddEndPuncttrue
\mciteSetBstMidEndSepPunct{\mcitedefaultmidpunct}
{\mcitedefaultendpunct}{\mcitedefaultseppunct}\relax
\EndOfBibitem
\bibitem{Wilks:1938dza}
S.~S. Wilks, \ifthenelse{\boolean{articletitles}}{\emph{{The large-sample distribution of the likelihood ratio for testing composite hypotheses}}, }{}\href{https://doi.org/10.1214/aoms/1177732360}{Ann.\ Math.\ Stat.\  \textbf{9} (1938) 60}\relax
\mciteBstWouldAddEndPuncttrue
\mciteSetBstMidEndSepPunct{\mcitedefaultmidpunct}
{\mcitedefaultendpunct}{\mcitedefaultseppunct}\relax
\EndOfBibitem
\bibitem{Feldman-Cousins}
G.~J. Feldman and R.~D. Cousins, \ifthenelse{\boolean{articletitles}}{\emph{Unified approach to the classical statistical analysis of small signals}, }{}\href{https://doi.org/10.1103/PhysRevD.57.3873}{Phys.\ Rev.\  \textbf{D57} (1998) 3873}, \href{http://arxiv.org/abs/physics/9711021}{{\normalfont\ttfamily arXiv:physics/9711021}}\relax
\mciteBstWouldAddEndPuncttrue
\mciteSetBstMidEndSepPunct{\mcitedefaultmidpunct}
{\mcitedefaultendpunct}{\mcitedefaultseppunct}\relax
\EndOfBibitem
\bibitem{cite:Plugin}
S.~Bodhisattva, M.~Walker, and M.~Woodroofe, \ifthenelse{\boolean{articletitles}}{\emph{{On the unified method with nuisance parameters}}, }{}Statist.\ Sinica \textbf{19} (2009) 301\relax
\mciteBstWouldAddEndPuncttrue
\mciteSetBstMidEndSepPunct{\mcitedefaultmidpunct}
{\mcitedefaultendpunct}{\mcitedefaultseppunct}\relax
\EndOfBibitem
\bibitem{LHCb-PAPER-2024-023}
LHCb collaboration, R.~Aaij {\em et~al.}, \ifthenelse{\boolean{articletitles}}{\emph{{Measurement of the CKM angle $\gamma$ in \mbox{$B^\pm\to DK^{*}(892)^{\pm}$} decays}}, }{}\href{https://doi.org/https://doi.org/10.1007/JHEP02(2025)113}{{JHEP} \textbf{02} (2025) 113}, \href{http://arxiv.org/abs/2410.21115}{{\normalfont\ttfamily arXiv:2410.21115}}\relax
\mciteBstWouldAddEndPuncttrue
\mciteSetBstMidEndSepPunct{\mcitedefaultmidpunct}
{\mcitedefaultendpunct}{\mcitedefaultseppunct}\relax
\EndOfBibitem
\bibitem{LHCb-PAPER-2023-029}
LHCb collaboration, R.~Aaij {\em et~al.}, \ifthenelse{\boolean{articletitles}}{\emph{{A model-independent measurement of the CKM angle $\gamma$ in partially reconstructed $\Bpm \to \Dstar h^\pm $ decays with $\D \to \KS h^{+}h^{-}$ ($h=\pi,K$)}}, }{}\href{https://doi.org/10.1007/JHEP02(2024)118}{JHEP \textbf{02} (2024) 118}, \href{http://arxiv.org/abs/2311.10434}{{\normalfont\ttfamily arXiv:2311.10434}}\relax
\mciteBstWouldAddEndPuncttrue
\mciteSetBstMidEndSepPunct{\mcitedefaultmidpunct}
{\mcitedefaultendpunct}{\mcitedefaultseppunct}\relax
\EndOfBibitem
\bibitem{LHCb-PAPER-2023-012}
LHCb collaboration, R.~Aaij {\em et~al.}, \ifthenelse{\boolean{articletitles}}{\emph{{Measurement of the CKM angle $\gamma$ using the $\Bpm\to \Dstar \hadron^\pm$ channels}}, }{}\href{https://doi.org/10.1007/JHEP12(2023)013}{JHEP \textbf{12} (2023) 013}, \href{http://arxiv.org/abs/2310.04277}{{\normalfont\ttfamily arXiv:2310.04277}}\relax
\mciteBstWouldAddEndPuncttrue
\mciteSetBstMidEndSepPunct{\mcitedefaultmidpunct}
{\mcitedefaultendpunct}{\mcitedefaultseppunct}\relax
\EndOfBibitem
\bibitem{LHCb-PAPER-2023-009}
LHCb collaboration, R.~Aaij {\em et~al.}, \ifthenelse{\boolean{articletitles}}{\emph{{Measurement of the CKM angle $\gamma$ in the \mbox{$\Bz \to \Dz \Kstarz$} channel using self-conjugate $\Dz \to \KS \hadron^+\hadron^-$ decays}}, }{}\href{https://doi.org/10.1140/epjc/s10052-023-12376-z}{Eur.\ Phys.\ J.\  \textbf{C84} (2024) 206}, \href{http://arxiv.org/abs/2309.05514}{{\normalfont\ttfamily arXiv:2309.05514}}\relax
\mciteBstWouldAddEndPuncttrue
\mciteSetBstMidEndSepPunct{\mcitedefaultmidpunct}
{\mcitedefaultendpunct}{\mcitedefaultseppunct}\relax
\EndOfBibitem
\end{mcitethebibliography}

\newpage
\centerline
{\large\bf LHCb collaboration}
\begin
{flushleft}
\small
R.~Aaij$^{38}$\lhcborcid{0000-0003-0533-1952},
A.S.W.~Abdelmotteleb$^{57}$\lhcborcid{0000-0001-7905-0542},
C.~Abellan~Beteta$^{51}$\lhcborcid{0009-0009-0869-6798},
F.~Abudin{\'e}n$^{57}$\lhcborcid{0000-0002-6737-3528},
T.~Ackernley$^{61}$\lhcborcid{0000-0002-5951-3498},
A. A. ~Adefisoye$^{69}$\lhcborcid{0000-0003-2448-1550},
B.~Adeva$^{47}$\lhcborcid{0000-0001-9756-3712},
M.~Adinolfi$^{55}$\lhcborcid{0000-0002-1326-1264},
P.~Adlarson$^{84}$\lhcborcid{0000-0001-6280-3851},
C.~Agapopoulou$^{14}$\lhcborcid{0000-0002-2368-0147}
% ...

{\footnotesize \it

% ...
$^{14}$Universit{\'e} Paris-Saclay, CNRS/IN2P3, IJCLab, Orsay, France\\
% ...
$^{38}$Nikhef National Institute for Subatomic Physics, Amsterdam, Netherlands\\
% ...
$^{47}$Instituto Galego de F{\'\i}sica de Altas Enerx{\'\i}as (IGFAE), Universidade de Santiago de Compostela, Santiago de Compostela, Spain\\
% ...
$^{51}$Physik-Institut, Universit{\"a}t Z{\"u}rich, Z{\"u}rich, Switzerland\\
% ...
$^{55}$H.H. Wills Physics Laboratory, University of Bristol, Bristol, United Kingdom\\
% ...
$^{57}$Department of Physics, University of Warwick, Coventry, United Kingdom\\
% ...
$^{60}$School of Physics and Astronomy, University of Glasgow, Glasgow, United Kingdom\\
$^{61}$Oliver Lodge Laboratory, University of Liverpool, Liverpool, United Kingdom\\
% ...
$^{69}$Syracuse University, Syracuse, NY, United States\\
% ...
$^{84}$Department of Physics and Astronomy, Uppsala University, Uppsala, Sweden, associated to $^{60}$\\
% ...
}
\end{flushleft}

\end{document}